\documentclass[onecolumn,conference,10pt]{IEEEtran}
\newcommand{\vect}[1]{{\mathbf{#1}}}
\newcommand{\norm}[1]{\| #1 \|}
\def \arg{\hbox{arg\ }}

\def\booknames#1/#2/#3/#4{#1:#2:#3:#4}
\usepackage{amsmath}
\usepackage[nothing]{algorithm}
\usepackage{algorithmic}
\usepackage{stfloats}
\usepackage{graphicx}
\usepackage{epsfig}
\usepackage{setspace}
\usepackage{color}

\usepackage[nomarkers]{endfloat}

\graphicspath{ {figures/} }

\usepackage{svn}
\SVN $Author: diego $
\SVN $Date$
\SVN $Rev: 398 $

\date{Rev. \SVNRev\ on \SVNDate\ --- \SVNAuthor}
\begin{document}

\doublespacing
%
\title{A Dual-based Method for Resource Allocation in
OFDMA-SDMA Systems with Minimum Rate Constraints}
%
\author{\IEEEauthorblockN{Diego Perea-Vega, Andr\'e Girard and Jean-Fran\c{c}ois Frigon (Corresponding author)}
\IEEEauthorblockA{\\Department of Electrical Engineering\\
{\'E}cole Polytechnique de Montr{\'e}al\\
C.P. 6079, succ. centre-ville, Montr{\'e}al, QC, Canada, H3C 3A7\\
Email: enrique.perea@polymtl.ca, andre.girard@gerad.ca, j-f.frigon@polymtl.ca\\
Tel: 1-514-340-4711 ext. 3642\\
Fax: 1-514-340-5892\\
}}

\maketitle
%
\begin{abstract}
We consider multi-antenna base stations using orthogonal
frequency-division multiple access (OFDMA) and space division
multiple access (SDMA) techniques to serve single antenna users,
where some of those users have minimum rate requirements and must
be served in the current time slot (real time users), while others
do not have strict timing constraints (non real time users) and
are served on a best effort basis. The resource allocation problem
is to find the user assignment to subcarriers and the transmit
beamforming vectors that maximize a linear utility function of the
user rates subject to power and minimum rate constraints. The
exact optimal solution to this problem can not be reasonably
obtained for practical parameters values of the communication
system. We thus derive a dual problem formulation whose optimal
solution provides an upper bound to all feasible solutions and can
be used to benchmark the performance of any heuristic method used
to solve this problem. We also derive from this dual optimal
solution a primal-feasible dual-based method to solve the problem
and we compare its performance and computation time against a
standard weight adjustment method. We find that our method follows
the dual optimal bound more closely than the weight adjustment
method. This off-line algorithm can serve as the basis to develop
more efficient
heuristic methods.
\end{abstract}

%
\section {Introduction} \label{sec:Introduction}

Multi-antenna base stations using orthogonal frequency-division
multiple access (OFDMA) and space division multiple access (SDMA)
can simultaneously transmit to different sets of users on multiple
subcarriers. In OFDMA-SDMA systems, multi-user diversity allows an
increase in the system throughput by assigning transmitting
resources to users with good channel conditions. High data rates
are thus made possible by exploiting the degrees of freedom of the
system in time, frequency and space dimensions. OFDMA-SDMA is also
supported by WiMAX and LTE-Advanced systems which are the
technologies that will most likely be used  to implement fourth
generation (4G) cellular networks~\cite{lte09a,wim09a}.

Due to the increased degrees of freedom
it is critical to use a  dynamic and efficient resource allocation
(RA) mechanism that takes full advantage of all OFDMA-SDMA transmitting
resources~\cite{letaief06}.
The role of an RA and scheduling algorithm is to allocate the
resources for transmission required to meet the quality of service
(QoS) requested from upper layers.
In this paper,  we focus on resource allocation policy for an
OFDMA-SDMA system supporting real time traffic with minimum rate
requirements.

%
%
\subsection{ State of the Art}
\label{sec:state_art}

The combinatorial nature of the RA problem makes it
NP-complete~\cite{bar07}. For an OFDMA-SDMA system with a
practical number of subcarriers, users and transmit antennas, it
is thus almost impossible to solve the RA problem directly.
Therefore, most research work focuses on  developing heuristic and
near-optimal RA algorithms.

Traffic in the system can be divided into two main groups: delay
sensitive real time (RT) services and delay-insensitive non real
time (nRT) services. Early work on OFDMA-SDMA systems focused on
solving the RA problem for only nRT services,where the objective
was to maximize the total throughput with only power constraints
and possibly minimum BER constraints. In~\cite{bar07} the complete
optimization problem was divided into a per-subcarrier user
selection problem and a power allocation problem, which are both
solved heuristically.
A similar approach using Zero Forcing (ZF) beamforming was
reported in \cite{ maciel07}. The  work of~\cite{bar07,maciel07}
does not solve the complete optimization problem because of its
computational complexity. Instead, it is separated into uncoupled
subproblems that provide non-optimal solutions. For this reason,
optimal or good approximations to near-optimal solutions are
important to benchmark the heuristic algorithm performance.
Several methods to obtain near optimum solutions have been
proposed for the RA problem with nRT traffic only. For example, in
\cite{ozbek09}  genetic algorithms are proposed, while
\cite{tsang04,chan07,xingmin10,perea10} provide methods to compute
a near-optimal solution based on dual decomposition methods. In
addition to providing a benchmark, near optimal algorithms can
also lead to the design of efficient RA methods as shown
in~\cite{perea10} where heuristic algorithms derived from the dual
decomposition methods are proposed.
%
%
%

Several approaches have been proposed to solve the OFDMA-SDMA RA
algorithm with RT traffic services. In some works, the design of
the RA algorithm supporting real time QoS is intertwined with the
scheduler design.  The scheduler decides at each time slot which
users must be served and sets their priorities in the RA utility
function~\cite{lee05}. Priority can thus be given according to the
current deadlines for RT services by increasing their weight in
the utility function while achieving some degree of fairness for
nRT services~\cite{song05}. The utility function is used as the
objective to maximize by the RA problem without any explicit
minimum rate constraints for the RT users. Similarly, RA heuristic
algorithms for RT services have also been proposed where the users
are served in priority according to the packet waiting time and
urgency to be served~\cite{tsai08,chung09,huang08}. Note that with
these \emph{reactive} approaches, the minimum rate requirements
are not modelled as hard constraints in the per-slot RA problem.
Instead, RT users with poor channel conditions are backlogged
until their delay is close to the deadline and then the RA reacts
and allocates resources to them. This causes an increase in the
average delay and delay jitter, which is not suitable for RT
services.

Another approach is to use constraints on the average rate
delivered to a user~\cite{tralli11}. However, unlike the work
presented in~\cite{wang-giannakis08} where a near optimal solution
is provided for the single antenna OFDMA RA problem with average
rate constraint, the algorithm presented in~\cite{tralli11} is a
heuristic approximation. Note that with average rate constraints,
RT users tend to be served when they have good channel conditions
which can create unwanted delay violations and jitter.

In~\cite{zhu10}, an heuristic algorithm for the RA problem with
proportional rate constraints is proposed. Although rate
constraints are used on a per slot basis, the actual rate is
relative to the total rate and does not offer a guaranteed minimum
rate to RT users.

Heuristic algorithms have also been proposed for the RA problem
with minimum rate constraints

~\cite{koutsopoulos02,papoutsis10}. For example, in
\cite{koutsopoulos02} a subcarrier exchange heuristic is proposed
to satisfy the minimum rates.

\subsection{Paper Contribution and Organization}

To guarantee the required QoS for RT users it is preferable to
integrate their minimum rate requirements into the RA optimization
problem rather than using reactive methods without explicit rate
constraints. Some heuristics have been proposed for the OFDMA SDMA
RA problem with minimum rate requirements but no optimal or
near-optimal solution has been derived, as was done for the
OFDMA-SDMA RA problem for nRT traffic. Having such a solution is
important to benchmark heuristics and can also lead to the design
of more efficient heuristics. The main contribution of this paper
is therefore an efficient method that provides a near-optimal
solution to the following OFDMA-SDMA RA problem for mixed RT and
nRT traffic:  for a given time slot, find the user selection and
beamforming vectors that maximize a linear utility function of the
users rates, given a power constraint and minimum rate constraints
for RT users. The user weights in the linear utility function are
arbitrary and can be the result of a prioritization or fairness
policy by the scheduler. We focus on the solution of the RA
optimization problem  by using a Lagrange dual decomposition
method. The solution to the dual function provides an upper bound
to the primal RA problem. We show, for small cases where it is
possible to find the optimal solution to the primal problem, that
 the duality gap is small. We also propose a simple
off-line near optimal algorithm which, based on the solution
obtained from the dual decomposition method, provides a feasible
solution to the RA problem. We study several cases where we
compare the performance of the dual upper bound, the feasible
solution provided by the proposed method and a solution obtained
using weight adjustments in the utility function. The results
indicate that the proposed method is close to the upper bound
while methods adjusting the user weights in the utility function
in order to prioritize RT users lead to significantly sub-optimal
solutions for the OFDMA-SDMA RA problem.

%
%

The paper is organized as follows. We describe the system and
formulate the optimization problem we seek to solve in
Section~\ref{sec:SysDescription}.  We present  in
Section~\ref{sec:dual-based-solution} the dual-based method. In
Section~\ref{sec:using-dual-method}, we present an algorithm that
finds a feasible solution based on the dual optimal solution and
we present two other alternative methods: one that solves the
problem by finding the exact solution, the other by using the
weight adjustment technique.  In Section~\ref{sec:results}, we
compare the performance and computation time of our method against
the two alterative methods and in Section \ref{sec:conclusions} we
present our conclusions.
%
\section {System description and Problem Formulation}
\label{sec:SysDescription}
We consider the resource allocation problem for the downlink transmission in a multi-carrier multi-user
multiple input single output (MISO) system with a single base station (BS) serving $K$ users, where some of those users have RT traffic with minimum rate requirements while the others have nRT traffic. The
base station is equipped with $M$ transmit antennas and each user
has one receive antenna. In this configuration, the BS can
transmit in the downlink data to different users on each subcarrier by performing
linear beamforming precoding at the BS. At each OFDM symbol, the
base station can change the beamforming vector for each user on
each subcarrier to maximize some performance function. In this paper, we assume that capacity achieving
channel coding is employed and the data rate units are in terms of bits per OFDM symbol or equivalently bits per second per Hertz (bps/Hz).

\subsection{Signal Model} \label{subsec:SignalModel}
First we describe the model used to compute the bit rate received by
each user. Define
\begin{description}
\item[$s_{k,n}$] the symbol transmitted by user $k$ on subcarrier
  $n$. We assume that the $s_{k,n}$ are independently identically
  distributed random variables
with $s_{k,n} \sim \mathcal{CN}(0,1)$.
\item[$\vect{w}_{k,n}$] a $M$-component column vector representing the
beamforming for user $k$ on subcarrier $n$. Unless otherwise
noted, we denote $\vect{w}$ the vector made up by the column stacking of the vectors
 $\vect{w}_{k,n}$.
\item[$\mathbf{x}_{n}$] a $M$-component column vector representing the
signal sent by the array of $M$ antennas at the BS  for each
subcarrier $n$.
\item[$\mathbf{h}_{k,n}$] a $M$-component row vector representing the
channel between the $M$ antennas at the BS and the receive antenna at user $k$  for each
subcarrier $n$.
\item[$z_{k,n}$] the additive white gaussian noise at the receiver for
  user $k$ on subcarrier $n$. The $z_{k,n}$ are independently identically distributed (i.i.d.) and without loss of generality we assume that $z_{k,n} \sim \mathcal{CN}(0,1)$.
\item[$y_{k,n}$] the signal received by user $k$ on subcarrier $n$.
\item[$r_{k,n}^0$] the rate of user $k$ on subcarrier $n$.
\end{description}
The signal vector $\vect{x}_{n}$ is built by a linear
precoding scheme which is a linear
transformation of the information symbols ${s}_{k,n}$:
\begin{equation}
\label{eq:x_signal}
\vect{x}_{n}= \sum_{k} \vect{w}_{k,n} s_{k,n}.
\end{equation}
The signal received by user $k$ on subcarrier $n$ is then given by
\begin{align}
y_{k,n}&= \vect{h}_{k,n} \vect{x}_{n}+z_{k,n} \notag\\
&=\vect{h}_{k,n} \vect{w}_{k,n} s_{k,n}
+ \sum_{j \neq k} \vect{h}_{k,n} \vect{w}_{j,n} s_{j,n} + z_{k,n}.
\label{eq:y_k_rx}
\end{align}
The second and third terms in the right side of (\ref{eq:y_k_rx})
correspond to the interference and noise terms, respectively.
Since the signals and noise are Gaussian, the interference plus
noise term is also Gaussian and the data rate of user $k$
for subcarrier $n$ is given by the Shannon channel capacity for an additive white Gaussian noise channel:
\begin{equation}
\label{eq:bitrate} r_{k,n}^0\left(\vect{w}\right)= \log_2 \left( 1
+ \frac {\norm{\vect{h}_{k,n} \vect{w}_{k,n}}^2} {1 + \sum_{j \neq
k}
  \norm{\vect{h}_{k,n} \vect{w}_{j,n} }^2 } \right).
\end{equation}
%
\subsection{Rate Maximization Problem}
\label{sec:rate-maxim-probl}

The general rate maximization problem corresponding to the
OFDMA-SDMA RA problem with mixed RT and nRT traffic is to find a
set of beamforming vectors $\vect{w}_{k,n}$ that will maximize the
users weighted sum rate. This is limited by the total power
available for the transmission at the base station and some users
with real time QoS requirements must receive a minimum rate. More
precisely, we assume that we know
\begin{description}
\item[$K$] Number of users in the cell.
\item[$\mathcal{K}$] Set of users in the cell: $\{1, \ldots, K\}$.
\item[$\mathcal{D}$]  A subset of $\mathcal{K}$ containing the
users that have minimum rate requirements. We define  $D= \vert
\mathcal{D} \vert $.
\item[$M$] Number of antennas at the BS.
\item[$\check{d}_{k}$] Minimum rate requirement for user
  $k$.
 \item[$N$] Number of subcarriers available.
\item[$\check{P}$] Total power available at the base station for transmitting over
all channels.
\item[$c_k$] Weight factors that are used by the scheduler to
implement prioritization or fairness.
\end{description}
We then want to solve the following optimization problem to obtain the resource allocation
%
\begin{align}
\label{eq:gral_form}
\max_{ \vect{w} } U^0 & = \sum_{n=1,k=1}^{N,K}
c_k r_{k,n}^0 ( \vect{w} )\\
\sum_{n=1,k=1}^{N,K} \norm{ {\vect{w}}_{k,n} }^2 & \leq
\check{P} \label{eq:powergen} \\
\sum_{n=1}^{N} r_{k,n}^0 ( \vect{w} ) & \ge \check{d}_{k} , \quad
k \in { \mathcal{D} }. \label{eq:minrategen}
\end{align}
The power used by the transmitter is represented by the sum of
the squared norms of the beamforming vectors in
constraint~(\ref{eq:powergen}). The achievable rate over all
subcarriers should be higher or equal than the required minimum
rate per user as in~(\ref{eq:minrategen}).

Problem~(\ref{eq:gral_form}-\ref{eq:minrategen}) is a non-convex,
nonlinear optimization problem. Using an exact algorithm to find a
global optimal solution is very hard considering the size of a
typical problem where there can be up to a hundred users and
hundreds of sub-channels. Another option is to use a standard
non-linear program (NLP) solver to compute a local optimal
solution and use different starting points in the hope of finding
a good global solution. The problem with this approach is that 1)
we don't know how close we are to the true optimum and 2) the
technique is quite time-consuming. Albeit those problems, we explored this
approach and observe
that most users end up with a zero
beamforming vector and only a small subset of users ($\leq M$)
actually get some rate. Furthermore, in accordance to what was reported for the SDMA problem in~\cite{yoo06}, we observed that at high SNR the ZF solution
is very close to the local optimum. This ZF solution is easily
computed by channel diagonalization and water-filling power
allocation. For these reasons, we now turn to the so-called
\emph{Zero-Forcing} beamforming strategy.
%
\subsection{Zero-Forcing Beamforming}
\label{subsec:ZF_stra}
In general, user
$k$ is subject to the interference from other users which reduces its
bit rate, as indicated by the denominator
in~(\ref{eq:bitrate}).  Zero-forcing beamforming is a strategy that
completely eliminates interference from other users.  For each
subcarrier $n$, we choose a set $\phi$ of $g \le M$ users which are
allowed to transmit. This is called a \emph{SDMA} set. We then impose
the condition that for each user $k$ in this set,
the beamforming vector of user $k$ must be orthogonal to
the channel vectors of all the other users of the set. This amounts to
adding the
orthogonality constraints
\begin{equation}
\label{eq:ortho_const} \vect{h}_{k,n} \vect{w}_{j,n} = 0 \quad j
\not= k,\, \, j, k \in \phi
\end{equation}
and the user $k$ data rate for subcarrier $n$ simplifies to:
\begin{equation}
\label{eq:zfbitrate} r_{k,n}^0\left(\vect{w}_{k,n}\right)= \log_2 \left( 1
+ \norm{\vect{h}_{k,n} \vect{w}_{k,n}}^2 \right).
\end{equation}
With zero-forcing, the beamforming problem is now made up of two
parts. We need to select a SDMA set for each subcarrier and for
each selected SDMA set, we must compute the beamforming vectors
in such a way that the total rate received by all users is
maximized. Because of this, we need to add another set of decision
variables
\begin{description}
\item[$\alpha_{k,n}$]  a binary variable that is 1 if we allow user
  $k$ to transmit on subcarrier $n$ and zero otherwise. We denote the collection of
  $ \alpha_{k,n} $ by
  the vector $\boldsymbol{\alpha}$ .
\end{description}
This results in the ZF problem
%
\begin{align}
\label{eq:objzf}
\max_{ \vect{w}, \boldsymbol{\alpha} } U^1 & = \sum_{n=1,k=1}^{N,K}
c_k r_{k,n}^0 ( \vect{w}_{k,n} )\\
\sum_{n=1,k=1}^{N,K} \norm{ \vect{w}_{k,n} }^2 & \leq
\check{P} \label{eq:powerz} \\
\sum_{n=1}^{N} r_{k,n}^0 ( \vect{w}_{k,n} ) & \ge \check{d}_{k} ,
\quad k \in { \mathcal{D} } \label{eq:minratezf} \\
\sum_k \alpha_{k,n} & \le M,  \quad \forall n \label{eq:sdmasize} \\
\left( \vect{h}_{k,n} \vect{w}_{j,n} \right)^2 & \le B \left[ ( 1
- \alpha_{k,n}) +  ( 1 - \alpha_{j,n} ) \right],  \quad \forall
n,\,
\forall k,\, \forall j, \, k \not= j \label{eq:zfconst} \\
\norm{\vect{w}_{k,n}} & \le A \alpha_{k,n} \label{eq:zfub}\\
\alpha_{k,n} \in \left\{ 0, 1 \right\} \label{eq:alphabin}
\end{align}
%
where $A$ and $B$ are some large positive constants.
Constraint~(\ref{eq:sdmasize}) guarantees that we do not choose
more than $M$ users for each subcarrier and
constraint~(\ref{eq:zfconst}) guarantees that if we have chosen
two users $k$ and $j$, they meet the zero-forcing constraints  and
that the constraints are redundant for other choices of users.
Constraint~(\ref{eq:zfub}) guarantees that the beamforming vector
is zero for users that are not chosen. It would seem that the
zero-forcing model is not improving things much: We have gone from
a non-convex nonlinear program to a non-convex mixed nonlinear
program. However, as we will explain in the next section, this allows us to design an efficient
and accurate algorithm.
\section{Dual-Based Solution Method}
\label{sec:dual-based-solution}
We cannot solve problem~(\ref{eq:objzf}--\ref{eq:alphabin}) fast
enough to use it  for a real time algorithm. Nevertheless, we need
to compute solutions so that we can use them as benchmarks to
evaluate the quality of real time heuristic approximations. We now
present an off-line solution technique that is tractable for
problems of moderate size and that can give either near-optimal
solutions or at least a bound when the optimum cannot be reached.

Solving the zero-forcing problem will require some form of search over
the $\alpha$ variables. Note that this ranges over all subsets of
users smaller than $M \times N$  so that the search
space is going to be fairly large. Our first transformation is thus to
separate the problem into single-subcarrier subproblems. For this, we
dualize the constraints~(\ref{eq:powerz}) and~(\ref{eq:minratezf}) since they are the ones that
couple the subcarriers. Define the dual variables
\begin{description}
\item[$\lambda$]  Lagrange multiplier for power constraint~(\ref{eq:powerz}).
\item[$\mu_k $]  Lagrange multipliers for minimum rate
  constraint~(\ref{eq:minratezf}) of   user $k$. The collection of
  $\mu_k$ is denoted $\boldsymbol{\mu}$.
\end{description}
In order to simplify the derivation,
we define the dual variables $\mu_k$ for all users $k \in
\mathcal{K}$. For users with no minimum rate requirements ($k
\notin \mathcal{D}$), we have $\mu_k = 0 $.
In what follows, we use the standard form of Lagrangian duality
which is expressed in terms of minimization  with inequality
constraints of the form $\le$. Under these conditions, the
multipliers $\lambda, \boldsymbol{\mu} \ge 0$. We get the partial
Lagrangian
\begin{align}
  \mathcal{L} & = - \sum_{n=1,k=1}^{N,K} c_k r_{k,n}^0 ( \vect{w}_{k,n}  )
  + \lambda \left[  \sum_{n=1,k=1}^{N,K} \norm{ \vect{w}_{k,n} }^2  - \check{P}  \right]
  + \sum_{k \in \mathcal{D} }  \mu_k \left[ \sum_{n=1}^{N}
    - r_{k,n}^0 ( \vect{w}_{k,n} )  + \check{d}_{k} \right]  \nonumber
  \\
  &= - \lambda  \check{P} + \sum_k   \mu_k \check{d}_{k} +
  \sum_n \left\{  - \sum_k  (c_k+\mu_k) r_{k,n}^0 ( \vect{w}_{k,n}  )
    + \lambda \sum_k \norm{ \vect{w}_{k,n} }^2
\right\} .
  \label{eq:partiallagr}
\end{align}
The value of the dual function $\Theta$ at some point $(\lambda,
\boldsymbol{\mu})$ is obtained by minimizing the Lagrange function
over the primal variables
\begin{equation}
  \Theta(\lambda, \boldsymbol{\mu} ) = \min_{\vect{w},
    \boldsymbol{\alpha} } \mathcal{L} ( \lambda, \boldsymbol{\mu} ,
  \vect{w}, \boldsymbol{\alpha} )
\label{eq:defdualf}
\end{equation}
and the dual problem is
\begin{align}
  \label{eq:defdualobj}
  \max_{\lambda, \boldsymbol{\mu} } \ & \Theta(\lambda, \boldsymbol{\mu} ) \\
  \lambda, \boldsymbol{\mu} & \ge 0 \label{eq:dualconst}
\end{align}
which we can solve by the well known subgradient
algorithm~\cite{bertsekas03}. From now on, we concentrate on the
calculation of the subproblem~(\ref{eq:defdualf}).
\subsection{Subchannel Subproblem}
\label{sec:subchsubpr}
Because of the relaxation of the carriers coupling constraints,
the subproblems in~(\ref{eq:defdualf}) decouple by subcarrier
since the objective~(\ref{eq:partiallagr}) is separable in $n$ and
so are constraints~(\ref{eq:sdmasize}--\ref{eq:zfub}). Computing
the dual function then requires the solution of $N$ independent
subproblems. For subcarrier $n$, after dropping the $n$ subscript,
this has the form
\begin{align}
  \label{eq:objsublagr}
  \min_{\vect{w}, \boldsymbol{\alpha} }  f_n &= - \sum_k  ( c_k + \mu_k
  )  r_{k}^0 ( \vect{w}_{k}  )
    + \lambda \sum_k \norm{ \vect{w}_{k} }^2 \\
\sum_k \alpha_{k} & \le M,  \label{eq:sdmasizelagr} \\
\left( \vect{h}_{k} \vect{w}_{j} \right)^2 & \le B \left[ ( 1 -
\alpha_{k}) +  ( 1 - \alpha_{j} ) \right], \quad \forall k, \,
\forall j,
\, k \not= j \label{eq:zfconstlagr}\\
\norm{\vect{w}_{k}} & \le A \alpha_{k} \label{eq:zfublagr} \\
\alpha_{k} \in \left\{ 0, 1 \right\} \nonumber
\end{align}
Problem~(\ref{eq:objsublagr}--\ref{eq:zfublagr}) is still a mixed
NLP, albeit of a smaller size.
\subsection{SDMA Subproblem}
\label{sec:sdma-subproblem}
A simple solution procedure is to enumerate all possible choices
for $\alpha_{k,n}$ that meet constraint~(\ref{eq:sdmasize}). This
is called the \emph{extensive} formulation of the problem. Each
such choice defines a SDMA set which we will denote by $s$  and
$\kappa = \vert s \vert$.  For each $s$, we solve the optimal
beamforming problem
%
%
\begin{align}
\label{eq:theo1_2prove-b}
\max_{ \vect{w} }  f_{n,s} & = \sum_{k \in s}  c_k^{\prime} \log_2 \left( 1 +
( \vect{h}_{k} \vect{w}_{k} )^2 \right) - \lambda \norm{ \vect{w}_{k} }^2 \\
\vect{h}_{j} \vect{w}_{k} & = 0 \quad j \not= k \,\ j,k \in s
\label{eq:orthosubpr}
\end{align}
%
where $c'_k = c_k + \mu_k$.  Note that
constraint~(\ref{eq:sdmasizelagr})  is automatically satisfied by
the construction of $s$, constraint~(\ref{eq:zfublagr}) simply
drops out since $\vect{w}_k = 0$ for $ k \not\in s$ and
constraint~(\ref{eq:zfconstlagr}) remains only for $k \in s$, but
we write it as~(\ref{eq:orthosubpr}) because we are considering
only users that belong to SDMA set $s$.

This is certainly not a feasible real time algorithm, but for
realistic values of $K$, say around 100, and $M=4$, the number of
cases is still manageable. This can give near-optimal solutions
against which to compare heuristics. This is possible only if the
SDMA beamforming
sub-problem~(\ref{eq:theo1_2prove-b}--\ref{eq:orthosubpr})  can be
solved efficiently.
\subsection{Beamforming Subproblem}
\label{sec:beamf-subpr}
This is in fact the case because it separates into $\kappa$
independent problems, one for each user in the SDMA set. Here again we
drop the $k$ index to simplify the discussion. We have to compute
the beamforming vector $\vect{w}$ for a given user. For this user, we
know the set of channel vectors for the other members of $s$. We
denote these vectors by the $(\kappa - 1) \times M$ matrix $\vect{H}$. We
also denote the channel vector for the user under consideration by
$\vect{h}$. The problem is then
\begin{align}
  \label{eq:usersubpobj}
  \max_{\vect{w}} f_{n,s,k}  & = c'_k \log_2 \left( 1 + \left( \vect{h}
      \vect{w} \right)^2  \right) - \lambda \norm{\vect{w}}^2 \\
  \vect{H} \vect w & = 0 \label{eq:usersubpconst}
\end{align}
so that
for realistic values, this is a small nonlinear
program. There are $M$ variables and $\kappa - 1$ linear
constraints. It can be solved quickly by a number of
techniques. Still, the overall  computation load can be quite large. There
will be $\kappa$ such problems to solve for each SDMA set, and there
are $ S =  \sum_{i=1}^\kappa \binom{K}{i}$ such sets for each of the
$N$ subcarriers so that we have to solve the problem $\kappa \times S
\times N$ times and this for each iteration of the subgradient
algorithm. Clearly, any simplification of the beamforming subproblem
can reduce the overall computation time significantly.
\subsection{Approximate Solution of the Beamforming Problem}
\label{sec:appr-solut-beamf}
This can be done by the following construction. Instead of
searching in the whole orthogonal subspace  of $\vect{H}$ as defined
by~(\ref{eq:usersubpconst}), we pick a
direction vector in that subspace and search only on its support.
This will give a good approximation to the extent that the
direction vector is close to the optimal vector. The choice of
direction is motivated by the fact that the objective function
depends only on the product $\vect{h} \vect{w}$. We then introduce
a new independent variable
\begin{equation}
  \label{eq:defqsubpr}
  q = \vect{h} \vect{w}
\end{equation}
and because this variable is not independent of $\vect{w}$, we
add Eq.~(\ref{eq:defqsubpr}) as a constraint. We then get the
equivalent problem
\begin{align}
  \label{eq:usersubpobeq}
  \max_{\vect{w}, q } f & = c' \log_2 \left( 1 + q^2  \right) - \lambda \norm{\vect{w}}^2 \\
  \left( \vect{h} \vect{w} \right)   & = q \label{eq:qwconstsub} \\
\vect{H} \vect w & = 0 \label{eq:usersubpceq}
\end{align}
which we can rewrite in the standard form $\vect{G} \vect{w} = \vect{b}$
where the $\vect{G}$ matrix is the concatenation of $\vect{h}$ and $\vect{H}$
and $\vect{b}^T = [q, 0 ,0 \ldots 0]^T$.

Since we are proposing to transform the constrained optimization
over the $\kappa$ variables into an unconstrained optimization
over $q$ only, we must be able to express $\vect{w}$ as a function
of $q$. Since the linear system is under-determined, this is
obviously not unique. We use $\vect{G}^+$, the pseudo-inverse of $\vect{G}$, for
this back transformation. We then have $\vect{w} = \vect{G}^+ \vect{b}$.
A well known property of the pseudo-inverse is that it picks the
vector of minimum norm compatible with the linear system. In other
words, choosing this transformation will \emph{minimize} $\vert
\vect{w} \vert$ so that it is minimizing the power term in the
objective function. Because $\lambda \ge 0$, this has the effect of
contributing to the maximization of $f$.

Expanding the matrix equation for $\vect{w}$, we find that
\begin{equation}
  \label{eq:wfp}
  w_i = \vect{G}^+_{i,1} q  \quad i = 1 \ldots \kappa
\end{equation}
which can then be replaced in the objective function. We get the
unconstrained problem
\begin{equation}
  \label{eq:unconstnotes}
  \max_q c'  \log_2 \left(1 + q^2 \right) - \lambda q^2  \gamma^2
\end{equation}
where $ \gamma = \norm{  \vect{G}^+_1} $ and $\vect{G}^+_1$ denotes the first
column of $\vect{G}^+$. Changing the variables $p = q^2$
and adding the constraint $p \ge 0$, we get the equivalent problem
\begin{align}
  \max_p  \ & c' \log (1 + p) - \lambda \gamma^2 p  \label{eq:notespinvobj}\\
  p & \ge 0
\end{align}
which has the solution
\begin{equation}
  \label{eq:1}
  p = \max \left\{  0,  \frac{c'}{\lambda \gamma^2} - 1 \right\}
\end{equation}
so that the computation time is basically the evaluation of $\vect{G}^+$.
\subsection{Computing the Dual Function}
\label{sec:comp-dual-funct}
To summarize, after reinstating all indices, we can write
%
\begin{align}
 \Theta(\lambda, \boldsymbol{\mu}) & = - \left( \lambda \check{P} - \sum_k \mu_k
  \check{d}_k + \sum_n f_n \right)   \\
f_n&=\max_s \left\{ f_{n,s}\right\} \label{eq:subproblem} \\
  f_{n,s} &= \sum_{k \in s} f_{n,s,k} \label{eq:fns-def} \\
  f_{n,s,k} & = c'_k \log \left( 1 + p_{n,s,k} \right) -
  \lambda \norm{\vect{w}_{n,s,k}}^2 \\
\label {eq:p-dual-def}
 p_{n,s,k} & =  \max \left\{ 0, \frac{c'_k}{\lambda
    \gamma_{n,s,k}^2} - 1\right\} \\
\label {eq:gamma-def}
\gamma_{n,s,k} & = \norm{\vect{G}_1^+}_{n,s,k}\\
\vect{w}_{n,s,k} & = \vect{G}^+_{n,s,k} \vect{b}_{n,s,k}\\
\vect{b}_{n,s,k} & = [p_{n,s,k}, 0, \ldots 0 ]^T
\end{align}
%
and $\vect{G}^+_{n,s,k}$ is the pseudo-inverse of the concatenation of
$\vect{h}_{n,k}$ and $\vect{H}_{n,s,k}$. We denote $s^*(n)$, $n=1,\dots,N$, the solution
of the maximization operation over $s$ in~(\ref{eq:subproblem}).
This is the optimal SDMA set
for subchannel $n$ for the current
values of the multipliers. We also denote $\vect{w}^*_{n,k}$ the
optimal beamforming vectors for the users $k \in s^*(n)$.
The largest part of the computation to evaluate the dual function
is the calculation of $\vect{G}^+_{n,s,k}$ which has to be done for each
subchannel, each SDMA set and each user in these SDMA sets. The number of
evaluations can become quite large but the size of each matrix is
relatively small so that the calculation remains feasible for
medium-size networks. Another advantage is that while solving the
dual problem requires multiple subgradient iterations, the
calculation of the pseudo-inverses is \emph{independent} of the
value of the multipliers. This means that can be done only
once in the initialization step of the subgradient procedure.

For
convenience we define $\Phi(\lambda, \boldsymbol{\mu}) $ as the
negative of the dual function,
%
\begin{align}
\label {eq:dual-redef}
 \Phi \doteq -\Theta = \lambda \check{P} - \sum_k \mu_k
  \check{d}_k +  \sum_n \max_s \left\{ f_{n,s}
  \right\}
\end{align}
and minimize this function when solving the dual problem
(\ref{eq:defdualobj}).
Algorithm~\ref{upper-bound-alg} finds the optimal dual variables
$(\lambda^*, \boldsymbol{\mu}^* )$ that solve the dual problem
(\ref{eq:defdualobj}) using the subgradient method~\cite{bertsekas03}
with a fixed step size $\delta$ . The optimum value $\Phi^*$ is a bound for the
primal objective and thus for any feasible point of the primal
problem (\ref{eq:objzf}). If
 $U^1$ is the objective achieved by any feasible point in the primal problem
(\ref{eq:objzf}) and $U^{*}$ its optimum, the following inequalities
hold \cite{boyd04}
\begin{equation}
  \label{eq:ineq-bound}
  \Phi(\lambda, \boldsymbol{\mu}) \geq \Phi^* \geq U^*  \geq U^1
\end{equation}
The dual optimum found $\Phi^*$, or any
approximation to it, is thus an \emph{upper bound} to the optimum value of the primal problem which can be used to benchmark other solution methods.
\begin{algorithm}
\begin{algorithmic}
\STATE Construct the set $\mathcal{S}$ of all subsets of users of size $1
\le \kappa \le M$
\FORALL { $n = 1 \ldots N$ } \FORALL {$s \in \mathcal{S}$}
\FORALL {$k \in s$}
\STATE Compute  the pseudo-inverse
$G^+_{n,s,k}$ and $\gamma_{n,s,k}$
\ENDFOR
 \ENDFOR
 \ENDFOR
 \STATE Choose an initial value $\lambda^{ 0 } $ and $\boldsymbol{\mu}^{
0  } $
\STATE Subgradient iterations. We set a limit of $I_m$ on
the number of iterations \FORALL {$i = 1 \ldots I_m$}
\STATE Solve the $N$ subproblems (\ref{eq:subproblem})
        \STATE Compute the subgradients:
        \STATE $g_{\mu}^{(k)} =  \check{d}_k -  \sum_{n} r_{n,k} $
        \STATE $g_{\lambda} = \sum_{n}  \sum_{k
          \in s^*(n) } \norm{\vect{w}^*_{n,k}  }^2 - \check{P} $
\IF {$\norm{\boldsymbol{g}_{\mu}} \le \epsilon $ and $\norm{g_{\lambda}} \le
\epsilon$}
\STATE Break
\ELSE
\STATE Update the multipliers
\STATE $\lambda^{i+1}= [\lambda^{i} + \delta  g_{\lambda} ]^+$
\STATE $\boldsymbol{\mu}^{i+1}= [\boldsymbol{\mu}^{i} + \delta \boldsymbol{g}_{\mu} ]^+$
\ENDIF
\ENDFOR
\end{algorithmic}
\caption{Calculation of the dual solution}
\label{upper-bound-alg}
\end{algorithm}
%
%

%
%

\subsection{Performance of the Dual Method}
\label{sec:computation-time}
\begin{figure}
  \centering
  \includegraphics[scale=0.75]{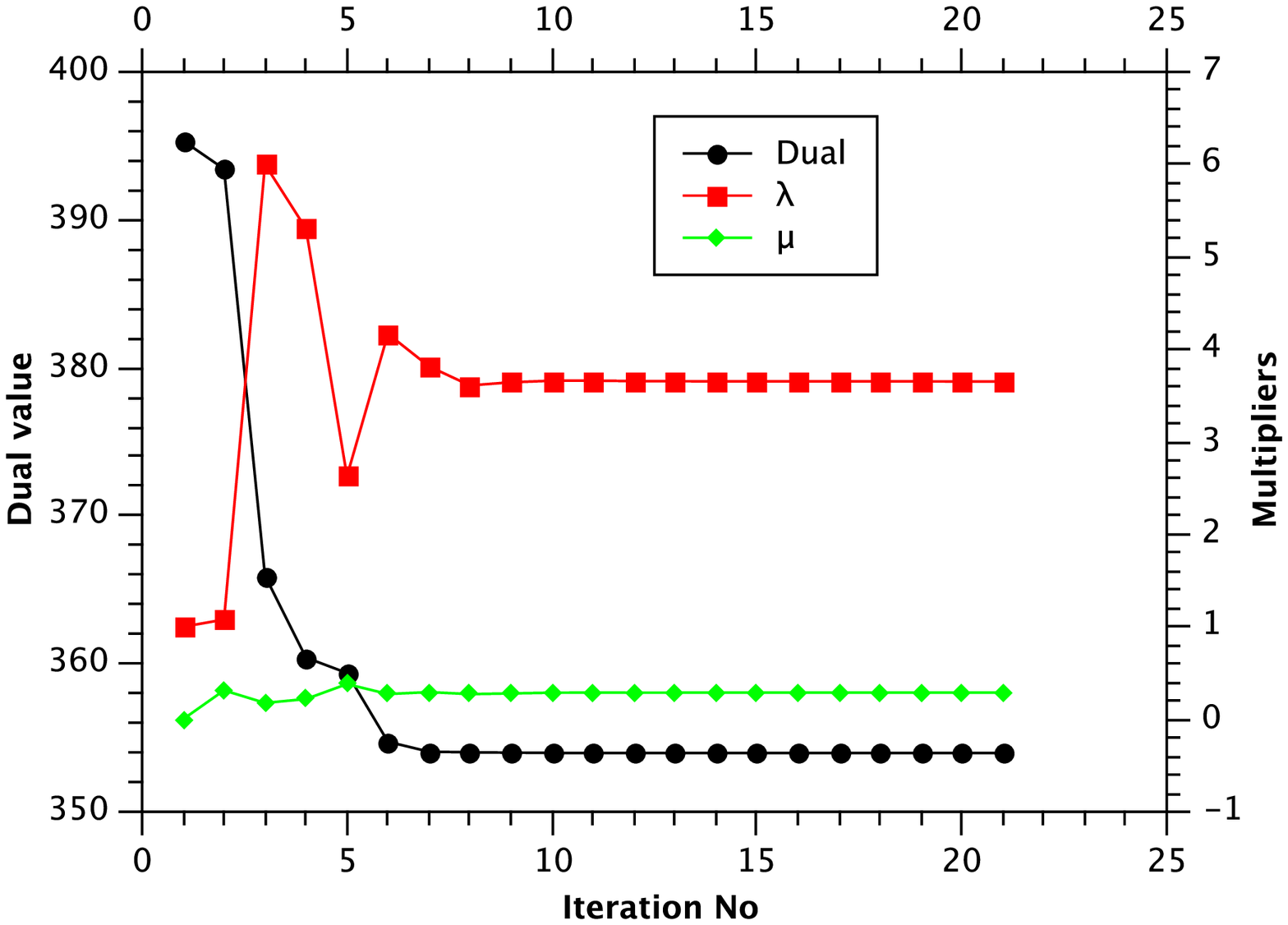}
  \caption{Dual function and multipliers for $M = 3$, $K = 16$, $N = 16$, $\check{P}=80$, $D=1$ and $\check{d}_1=$20 bps/Hz.}
  \label{fig:dual_min_iter}
\end{figure}

\begin{figure}
  \centering
  \includegraphics[scale=0.75]{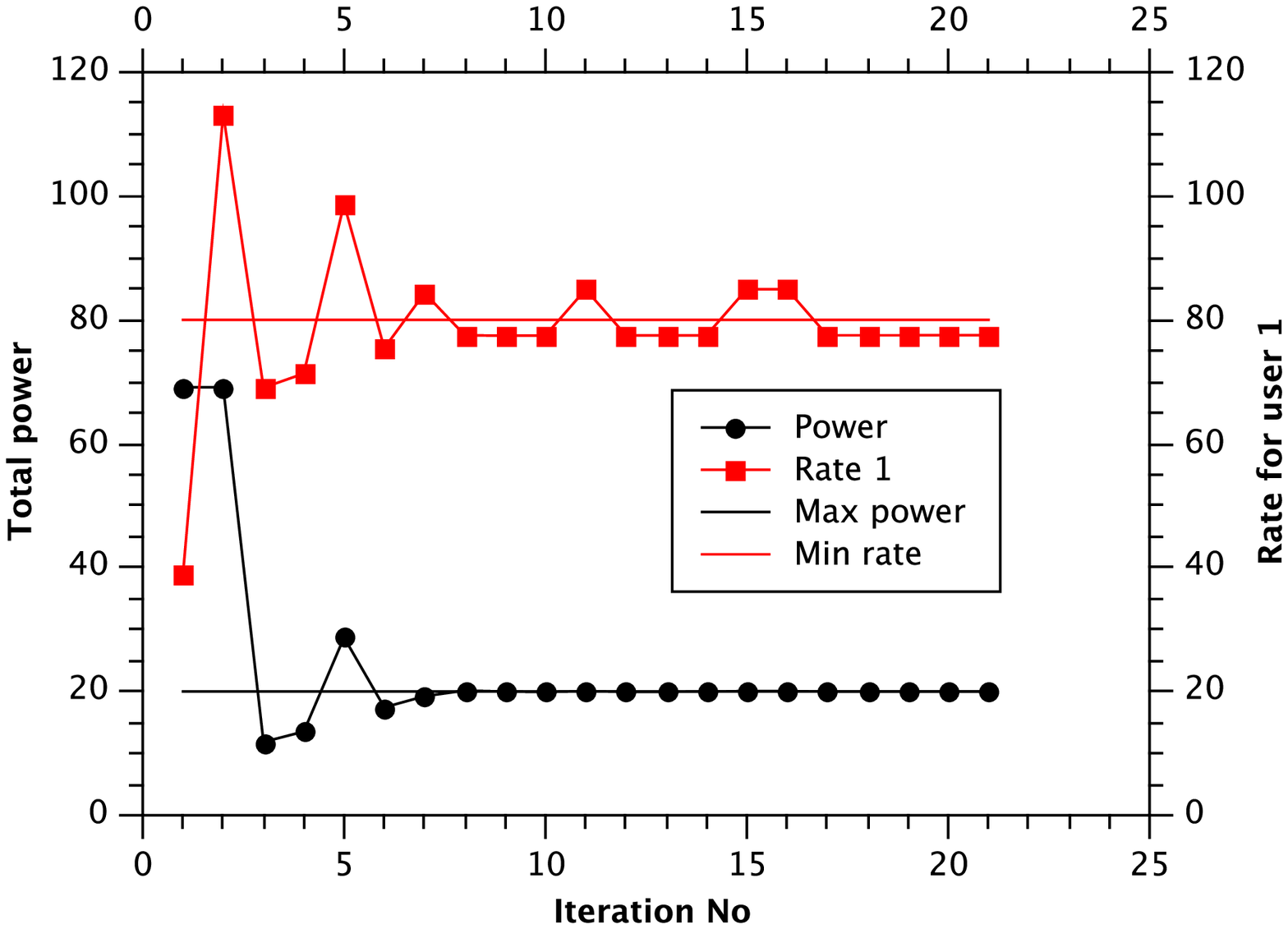}
  \caption{Power and rate constraints for $M = 3$, $K = 16$, $N = 16$, $\check{P}=80$, $D=1$ and $\check{d}_1=$20 bps/Hz.}
  \label{fig:dual_iter_feas}
\end{figure}

We study in this section the
convergence speed and computation time of the dual algorithm for single random channel realizations (the average performance is studied in Section~\ref{sec:results}). We present in
Figure~\ref{fig:dual_min_iter} the value of the
dual function and Lagrange multipliers as a function of the
number of iterations while the total transmit power and the rate received by user 1 (single user with minimum rate requirements) are shown in Figure~\ref{fig:dual_iter_feas}. We can see that the method
converges very quickly  to
a solution that is both close to the minimum value and feasible.
We observed a similar behavior for several other configurations.

\begin{figure}
  \centering
  \includegraphics[scale=0.75]{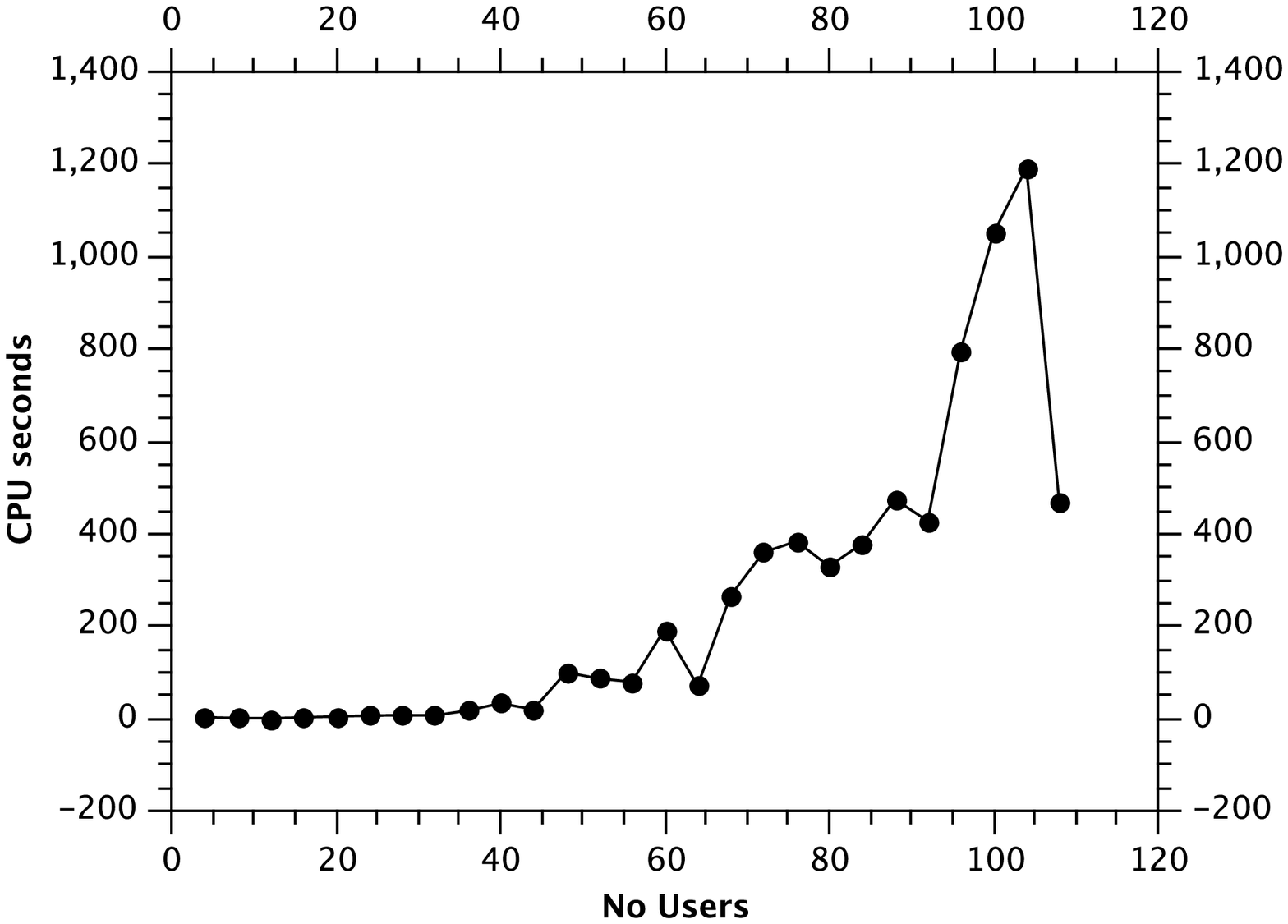}
  \caption{Computation time as a function of the number of users $K$ for $M = 3$, $N = 4$, $\check{P}=80$, $D=1$ and $\check{d}_1=$20 bps/Hz.}
  \label{fig:cpu_vs_k}
\end{figure}

\begin{figure}
  \centering
  \includegraphics[scale=0.75]{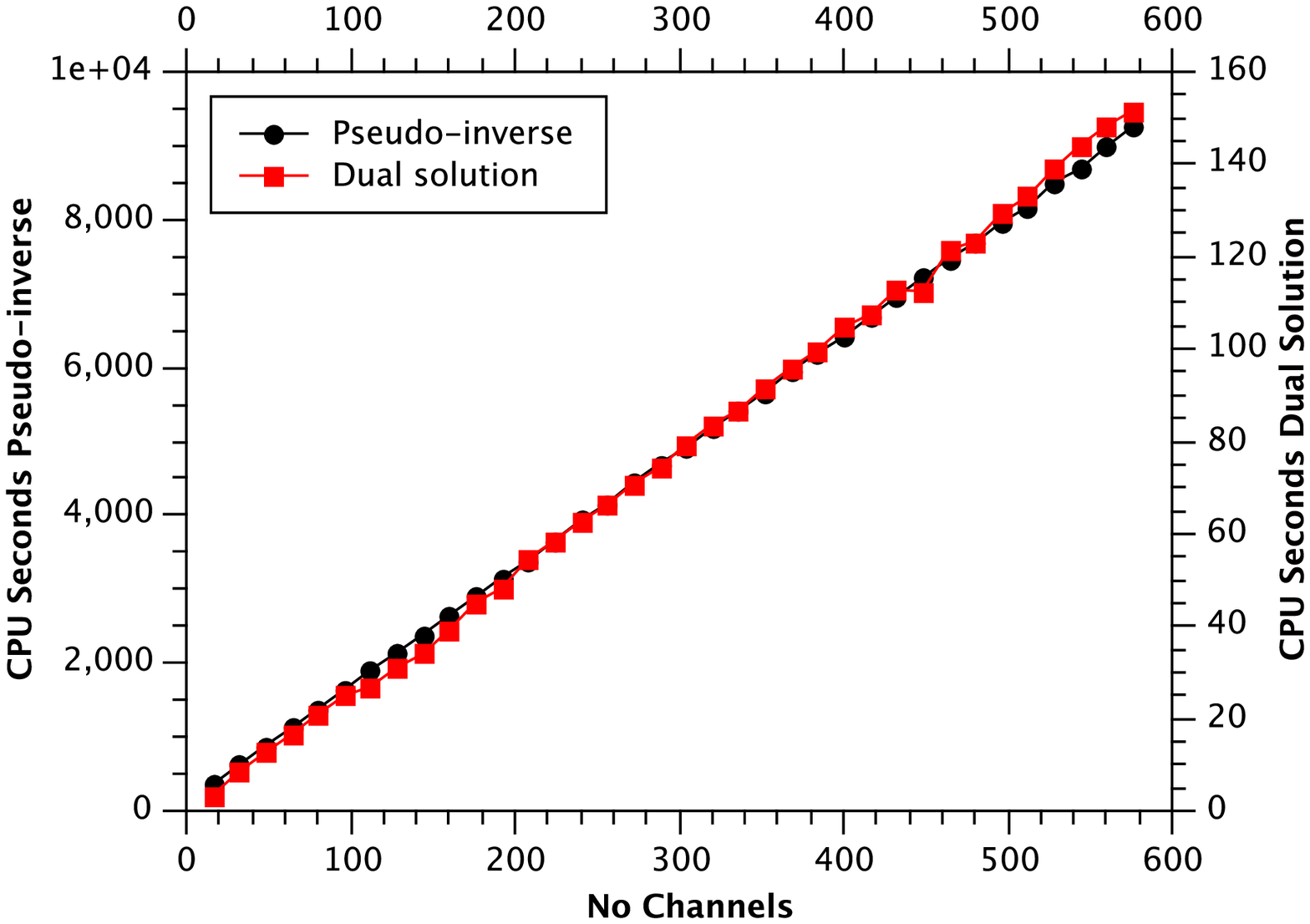}
  \caption{Computation time as a function of the number of subcarriers $N$ for $M = 3$, $K=16$, $\check{P}=80$, $D=1$ and $\check{d}_1=$20 bps/Hz.}
  \label{fig:cpu_vs_n}
\end{figure}

We can see the computation time (i.e., the time used by the CPU)
required to solve a problem as a function of the number of users
in Figure~\ref{fig:cpu_vs_k} and as a function of the number of
channels in Figure~\ref{fig:cpu_vs_n}. In
Figure~\ref{fig:cpu_vs_n} we further separated the CPU time
between the CPU time required to compute the pseudo-inverse at the
initialization and the CPU time required to find the dual
solution. There is a striking difference since the CPU time growth
is much faster than linear as a function of $K$ while it is very
linear as a function of $N$. This is obviously because the
Lagrangian decomposition separates the overall problem into $N$
independent subproblems and for fixed $K$, the CPU time will grow
as the number of subproblems. As a side note, the last value
plotted in Figure~\ref{fig:cpu_vs_k} is much lower because for
this random channel realization the problem had no active rate
constraints, thus no extra iterations were required.  Another
important point however is that calculating the pseudo-inverses is
much more time-consuming that solving the dual itself.  Notice
that the two curves in Figure~\ref{fig:cpu_vs_n} almost overlap
but they are plotted on a different scale.
%
%
%
\section{Other Solution Algorithms}
\label{sec:using-dual-method}
The SDMA set selection and beamforming vectors found by
algorithm~\ref{upper-bound-alg} do not always provide a primal
feasible solution.  The rate or power constraints might be violated
whenever the algorithm stops because the number of
iterations has been reached before the convergence
rule is met. In this section we present three different approaches to solve the primal problem.
In Section~\ref{sec:finding-an-exact}, a direct method where we enumerate all the
variables $\boldsymbol{ \alpha}$ and solve the problem for the
beamforming vectors $\vect{ w }$ in each case is explained. In Section~\ref{sec:find-feas-solut},
we propose a simple procedure to obtain a feasible solution from the dual solution found with algorithm~\ref{upper-bound-alg}. Finally, in Section~\ref{sec:weight-adjustm-meth} we propose a method based on weight adjustments of the utility function to meet the minimum rate requirements.
In Section~\ref{sec:difference-with-dual} we study
the difference between the dual-based and the weight adjustment methods.
\subsection{Exact Solution}
\label{sec:finding-an-exact}
One way to evaluate the accuracy of the dual algorithm is to
compare it with an exact solution.
Problem~(\ref{eq:objzf}--\ref{eq:zfub}) is a nonlinear MIP for
which we can do a complete enumeration of the binary variables
$\boldsymbol{\alpha}$, the set of $\vect{w}$ variables is
determined by the choice of   a particular $\boldsymbol{\alpha}$
and constraints~(\ref{eq:zfub}) are automatically satisfied. The
same remark goes for Eq.~(\ref{eq:zfconst}) where the only
remaining constraints are the ones for which $\alpha_{k,n} = 1$.
 The ZF constraints are now written as (\ref{eq:zfconst_copy})
 where $s(n)$ is the SDMA set for subchannel $n$, given by the value of
 $\vect{\alpha}$.
For each given value of $\boldsymbol{\alpha}$, we then  need to
solve the optimal beamforming problem
%
\begin{align}
\label{eq:fix_alpha_prob} \max_{ \vect{w} }  \sum_{n=1,k=1}^{N,K}
c_k r_{k,n}^0 ( \vect{w}_{k,n} ) \\
\sum_{n=1,k=1}^{N,K} \norm{ \vect{w}_{k,n} }^2 & \le  \check{P}
\label{eq:powerz_copy} \\
\sum_{n=1}^{N} r_{k,n}^0 ( \vect{w}_{k,n} ) & \ge  \check{d}_{k}
\quad k \in { \mathcal{D} } \label{eq:minratezf_copy} \\
\left( \vect{h}_{k,n} \vect{w}_{j,n} \right)^2 & = 0\quad \, j,k
\in s(n), \, \, k \not= j \, \, \forall n \label{eq:zfconst_copy}
\end{align}
where the optimization variables are the beamforming vectors of users in
$s(n)$ for each subchannel $n$.
The main difference with the dual
method is that here, we have to enumerate the set of all possible
choices of subsets of $K$ users, $M$ antennas and $N$ channels.
This is a much larger set than for the dual method where the
enumeration is done separately for each channel and for this
reason, we cannot expect to solve very large problems with the exact solution approach.

Each beamforming sub-problem~(\ref{eq:fix_alpha_prob}) is
relatively small but it is not convex. We  use the same technique
as in Section~\ref{sec:appr-solut-beamf} to simplify it. First,
we group the user vectors belonging to SDMA set $s(n)$ in a
$|s| \times M$ matrix $\vect{H}_n$ and we assume a given $1 \times
|s|$ user power vector $\vect{p}_n$.  The ZF constraints
(\ref{eq:zfconst_copy}) are written in matrix form as
%
\begin{align}
{\vect{H}_n} \vect{W}_n = \text{diag}( \sqrt{ \vect{p}_n } ),
\quad  \forall n \label{eq:zfconst_mat}
\end{align}
and  we restrict the beamforming vectors to be in the direction
given by the pseudo-inverse matrix
%
\begin{equation}
\label{eq:psudo-sol}
\vect{W}_n = {\vect{H}_n}^{\dag}
\text{diag}( { \sqrt{ \vect{p}_n } } ),   \quad  \forall n.
\end{equation}
%
The problem then reduces to the optimization over the vector
$\vect{p}_n$
%
 \begin{align}
 \label{eq:q_prob}
\max_{ \vect{ p}_n  } U^2 & = \sum_{k \in s(n)} c_k \sum_{n=1}^N \log_2 ( 1 + p_{k,n} )\\
 \sum_{n=1}^N \sum_{k \in s(n)} \beta_{k,n} p_{k,n}  & \le  \check{P}   \label{eq:q-powerz} \\
\sum_{n=1}^N \log_2 ( 1 + p_{k,n} )  & \le  \check{d}_{k},
\quad k \in { \mathcal{D} } \label{eq:q-minratezf} \\
\beta_{k,n} & =  {\left[ {(\vect{H}_n^{\dag})}^H
    \vect{H}_n^{\dag} \right]}_{k,k} \nonumber \\
\vect{p}_n & \ge 0. \label{eq:pgt0}
\end{align}
%
Problem (\ref{eq:q_prob}--\ref{eq:pgt0}) is convex since we are maximizing a
concave function over a convex set and can be solved by standard techniques.
The overall  procedure  to find an exact solution by enumeration
is summarized in algorithm~(\ref{algo:enum-alg}).
\begin{algorithm}
\begin{algorithmic}
\STATE { MAX $\leftarrow 0$ } \FOR { $i=1$ to $S^N$ }
    \STATE{ Solve problem (\ref{eq:q_prob}--\ref{eq:pgt0}) for given SDMA set
    assignment $s_i(n)$  with objective function $U^2$ }
    \IF { $ U^2 \ge $ MAX  }
        \STATE { MAX $\leftarrow U^2$ }
        \STATE { $s^o \leftarrow s_i$ }
    \ENDIF
\ENDFOR
\end{algorithmic}
\caption{Enumeration Algorithm}
\label{algo:enum-alg}
\end{algorithm}
\subsection{Dual-Based Feasible Solution}
\label{sec:find-feas-solut}
Algorithm~\ref{upper-bound-alg} can provide an optimal solution
$(\lambda^*,\boldsymbol{\mu}^*)$  to the dual
problem~(\ref{eq:defdualobj}).  However, the subgradient algorithm
is known to converge slowly and, in some cases, we need to stop
the iterations before the algorithm has reached an optimal dual
solution. In these cases, the optimal solution may not be
feasible. In this section we are proposing a refinement of the
dual method to construct a feasible solution starting from the
final solution found for the dual problem.

Algorithm~\ref{algo:dualfeas} summarizes this method. The
algorithm begins by solving the dual problem~(\ref{eq:defdualobj})
using algorithm \ref{upper-bound-alg}. If the solution is not
feasible either directly or by recomputing the  power allocation
using (\ref{eq:q_prob}--\ref{eq:pgt0}) for the SDMA set assignment
found in the dual problem, the algorithm performs a search by
increasing the dual variables associated to the users whose QoS
constraints are not met until a new SDMA set assignment is found.
It then solves the power allocation problem
(\ref{eq:q_prob}--\ref{eq:pgt0}) for this new SDMA set assignment
and checks the solution feasibility with regards to the minimum
rate constraints. The search for new SDMA sets continues using
this method until a feasible SDMA set assignment is found or a
maximum number of iteration is reached.




\begin{algorithm}
\begin{algorithmic}
\STATE { Solve the dual problem~(\ref{eq:defdualobj}) using
algorithm \ref{upper-bound-alg}.  This yields the optimal dual
variables $ \lambda^*, \boldsymbol{\mu}_k^*  $ and a SDMA set
assignment vector ${s}^*(n)$ for each subchannel $n$.}
\STATE{ Set $s^o_0(n) = s^*(n)$} \STATE {Evaluate total power and
user rate constraints (\ref{eq:powerz}--\ref{eq:minratezf})}
\IF{All constraints are met}
   \STATE { Exit. A feasible solution has been found.}
\ENDIF
\STATE {Compute power allocation problem~(\ref{eq:q_prob}--\ref{eq:pgt0}) for $s^o_0(n)$ and evaluate total power and user rate constraints (\ref{eq:powerz}--\ref{eq:minratezf}) }
\IF{All constraints are met}
   \STATE { Exit. A feasible solution has been found.}
\ENDIF

\STATE {Compute the multipliers $\mu_k$  for users $k$ such that $r_k < \check{d}_k$}
\FOR {  $j = 1$ to $\overline{J}$  }
       \STATE { $\mu_k = \mu_k + \delta$ }
       \STATE { Find $s^o_j= \arg \max_{s} \{ f_{n,s} \}$
        where $f_{n,s}$ is given by (\ref{eq:fns-def}) for the current dual
        variables $\lambda, \boldsymbol{\mu}$ }
        \STATE{ Let $s^o_j(n)$ be the SDMA assignment found}
        \IF { $s^o_j(n) \not= s^o_{j-1}(n)$}
              \STATE { We have found a new SDMA assignment}
              \STATE {Compute power allocation problem~(\ref{eq:q_prob}--\ref{eq:pgt0}) for $s^o_j(n)$ and evaluate total power and user rate constraints (\ref{eq:powerz}--\ref{eq:minratezf}) }
                 \IF{All constraints are met}
                    \STATE { Exit. A feasible solution has been found.}
                 \ENDIF

        \ENDIF
 \ENDFOR
 \STATE { Exit. A feasible solution was not found.}

\end{algorithmic}
\caption{Calculating a feasible point from the dual solution}
\label{algo:dualfeas}
\end{algorithm}
%
%
In contrast to the enumeration method described in
Section~\ref{sec:finding-an-exact} which performs an enumeration
of all possible SDMA set assignments, the dual-based
algorithm~\ref{algo:dualfeas} is a method that finds a SDMA set
assignment close to the dual optimal and then uses it to solve one
sub-problem (\ref{eq:q_prob}--\ref{eq:pgt0}). This makes the
search for a near-optimal feasible solution much faster than
finding the exact solution.
\subsection{Weight Adjustment Method}
\label{sec:weight-adjustm-meth}
In Section~\ref{sec:state_art} we discussed several RA algorithms
that support RT traffic which increase the user weights in the
utility function until such users receive transmission resources.
In this section we thus propose a weight adjustment method to
evaluate the efficiency of algorithms that use this approach. The
objective of the proposed method is to find a set of weights in
the utility function~(\ref{eq:objzf}) for which, when we solve
problem~(\ref{eq:objzf}--\ref{eq:zfub}) without the rate
constraints~(\ref{eq:minratezf}), the rate requirements of the RT
users are met. We also want the set of weights to have the least
deviation between users in order to maximize the multi-user
diversity gain.
Algorithm~\ref{subsubsec:ck-method-alg} implements a generic
method for weight adjustment to achieve this objective. The algorithm increases the user
weights for RT users until enough resources are allocated to meet the minimum rate requirements.  The
parameter $\epsilon$ controls how much the weights are increased
with respect to the rate bounds.
%
\begin{algorithm}
\begin{algorithmic}
   \STATE {Solve RA problem (\ref{eq:objzf}) without minimum rate
constraints constraints (\ref{eq:minratezf}) }
   \STATE { $\vect{c}' \leftarrow \vect{c} $ }
   \STATE { Let $r_k$ be the achieved rate for user $k$ at every
   iteration }
   \STATE { iteration $\leftarrow$ 1 }
   \WHILE { ($r_k < \check{d}_k$ for one or
more users $k \in {\mathcal{D}}$) AND (iteration $\leq $
MAX\_iterations) }
       \STATE {Increase user weight using $c_k^{\prime} = c_k^{\prime} +
\epsilon \left( \check{d}_k - r_k  \right) $ for users in need, where $0<
\epsilon \leq 1$  }
           \STATE { Solve RA problem (\ref{eq:objzf}) without minimum rate
           constraints (\ref{eq:minratezf}) using user weights {$\vect{c}'$}
                \STATE { iteration $\leftarrow$ iteration $+1$ }
} \ENDWHILE
\end{algorithmic}
\caption{Weight adjustment algorithm}
 \label{subsubsec:ck-method-alg}
\end{algorithm}
%
%
\subsection{Comparison of the Weight Adjustment and the Dual-Based Methods}
\label{sec:difference-with-dual}
The rates achieved by weight adjustment algorithm~\ref{subsubsec:ck-method-alg} and the
dual-based algorithm \ref{algo:dualfeas} are different
since they solve different problems.
That is, algorithm~\ref{subsubsec:ck-method-alg} can be seen as
solving problem~(\ref{eq:q_prob}--\ref{eq:pgt0}) by a linear penalty method for
constraints~(\ref{eq:q-minratezf}) of the form
\begin{displaymath}
  P_k = \min \left\{ 0, r_k - \check{d}_k \right\}
\end{displaymath}
The modified objective function is then
\begin{align}
  U_P & = \sum_k c_k r_k +  P_k \nonumber \\
  & = \sum_k c_k r_k + \epsilon \sum_{k \vert r_k < \check{d}} (  r_k - \check{d}
  )  \label{eq:penaltyobj}
\end{align}
At each iteration of the penalty method, whenever rate constraints
are active, the solution of~(\ref{eq:penaltyobj}) cannot be
smaller than that of~(\ref{eq:q_prob}--\ref{eq:pgt0}) since it is
a relaxation. Notice that problem (\ref{eq:penaltyobj}) is quite
simple since it has a single constraint~(\ref{eq:q-powerz}) but it
has to be solved many times to adjust the weights of the real time
users. In weight adjustment algorithms such as
\cite{tsai08,chung09}, the user weights are increased  at each
time slot using an increasing function of the packets delay, so
the computation task is distributed over time. However, this
distributed approach does not guarantee that the rate requirements
are met in a given time slot and leads to delay violations and
 jitter.

%
%

\begin{figure}
\centering
\includegraphics[scale=0.7]{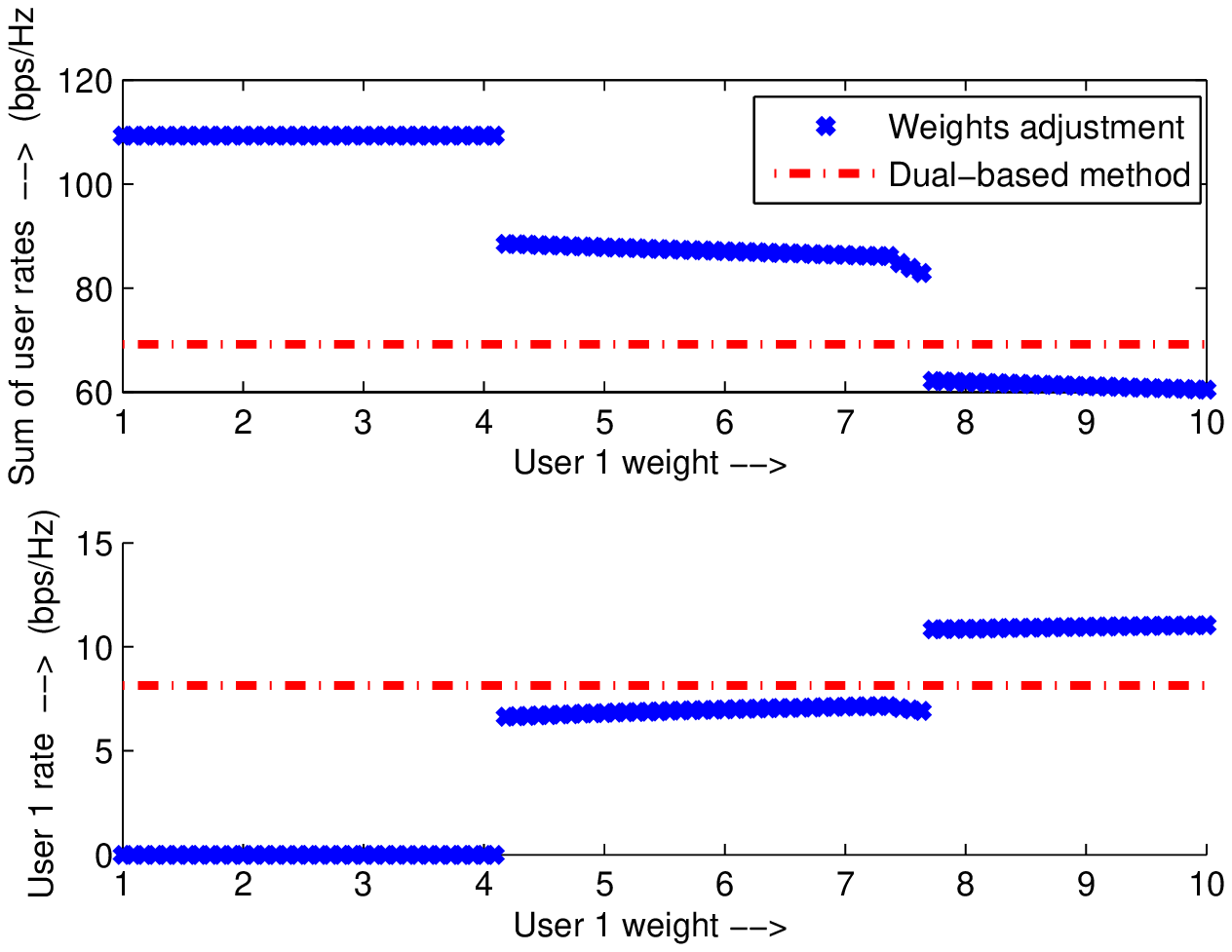}
\caption{Comparison between weight adjustment and dual-based methods for $M = 3$, $K=6$, $N=4$, and $\check{P}=20$.}
\label{fig:alg-ck-comparison2}
\end{figure}

To illustrate numerically the difference between solving the
problem with explicit rate constraints versus modifying the user
weights, we simulated the performance of both approaches for a
single random channel realizations. We set a minimum rate
constraint of $\check{d}_1= 8$ bps/Hz for one RT user.  We first
solved the problem using the dual-based
algorithm~\ref{algo:dualfeas} using equal weights for all users.
The sum rate is shown with the dash line in the top plot of
Figure~\ref{fig:alg-ck-comparison2} while the user 1 rate is shown
with the dash line in the bottom plot of
Figure~\ref{fig:alg-ck-comparison2}. To specifically study the
impact of the utility function weights on the performance, we then
solved problem~(\ref{eq:objzf}--\ref{eq:zfub}) without the rate
constraints~(\ref{eq:minratezf}) for different weight assignments
as follows: All user weights were set to 1 initially and we varied
the user 1 weight from 1 to 10. The corresponding user 1 and sum
rates are indicated in Figure~\ref{fig:alg-ck-comparison2} with
the star line.  We can observe  three regions as a function of the
user 1 weight $c_1$. For low values of $c_1$, user 1 does not get
the required minimum rate so that the total rate is much larger
that the optimal value, which is to be expected for an unfeasible
solution. For middle values, the user 1 rate is much better but is
not feasible and the total rate is correspondingly smaller but
still larger than the optimum. Finally, if we increase $c_1$ such
that the required rate must be satisfied to a high accuracy, the
total rate is lower than the value obtained with the dual-based
method. Note that the total rate and user 1 rate discontinuities
observed for the weight adjustment method are due to the fact that
the selected SDMA set changes as we increase the weight and the
solution moves from one region to another.

\begin{figure}
\centering
\includegraphics[scale=0.6]{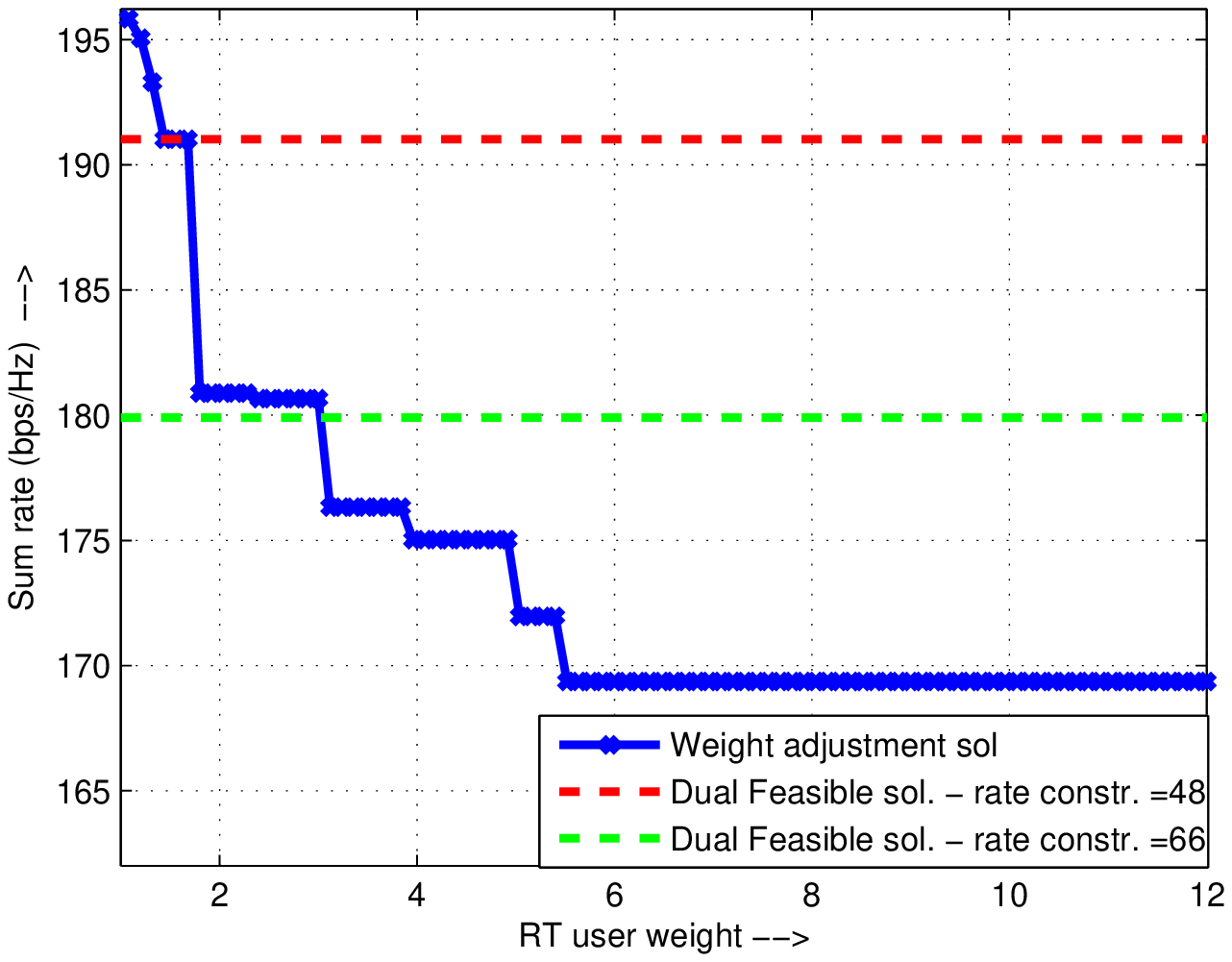}
\caption{Total rate for weight adjustment and dual-based methods for $M = 3$, $K= 8$, $N=8$, and $\check{P}= 20$.}
\label{fig:totalrate_weight_meth_combo2}
\end{figure}

\begin{figure}
\centering
\includegraphics[scale=0.6]{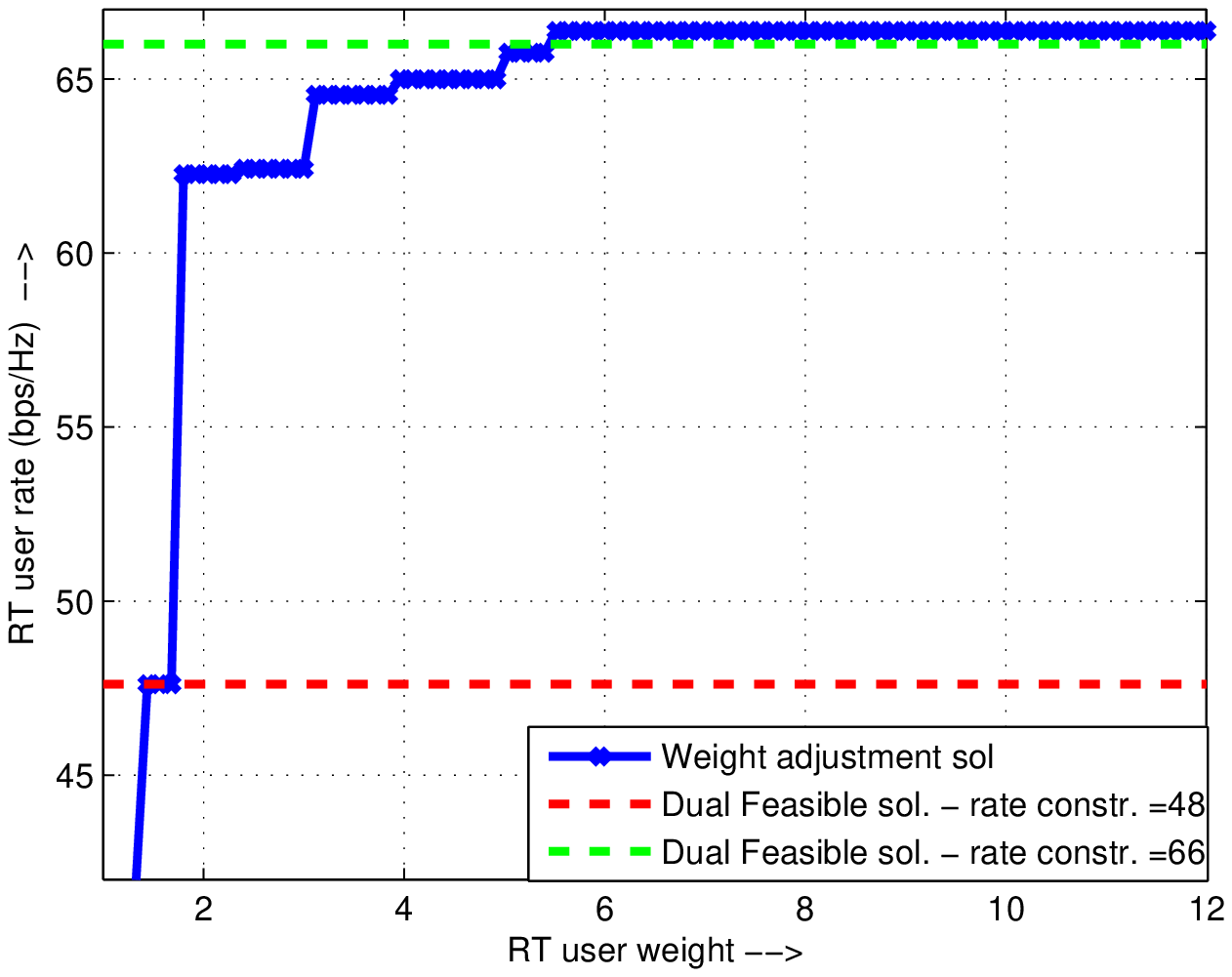}
\caption{User 1 rate for weight adjustment and dual-based methods for $M = 3$, $K= 8$, $N=8$, and $\check{P}= 20$.}
\label{fig:user1rate_weight_meth_combo2}
\end{figure}

Another example is shown in Figures~\ref{fig:totalrate_weight_meth_combo2} and~\ref{fig:user1rate_weight_meth_combo2} where we plot the total objective and the user rate, respectively, as a function
of $c_1$.  We consider two cases for the user 1 rate requirement, one where $\check{d}_1= 48$ bps/Hz
 and the other where $\check{d}_1= 66$ bps/Hz.

The results
show that for a value of $c_1\approx 1.5$ the weight adjustment approach provides the same total
rate as the dual-based method and the user 1 rate requirement is met.
However, it is interesting to note that the range of $c_1$ over which this is
possible is quite narrow: A slightly lower value produces an
unfeasible solution and a slightly higher value, a much larger rate
allocation for the user 1 with a lower total rate. In other
words, one would need many iterations of the weight adjustment
algorithm to get the required accuracy to find the optimum.
For the other case, where the rate constraint is higher, we see that the weight adjustment method is unable to
find a solution that is both feasible and reasonably close to the
optimal value. This is a clear indication that the rate adjustment
algorithm should be used with care, especially in those cases where the
rate constraints are important.

These examples show that even if we were to perform an exhaustive
search to determine the weights $\vect{c}'$, the performance would
not be as good as the dual-based method. For the cases where the
performance is comparable, we also observed that the weight
adjustment method performance is very sensitive to the weight
value, which is not a desirable property. However, if an efficient
method to adjust the weights is used, the computational load of
the weight adjustment method would be lower than that of the
proposed dual-based method because it solves a problem with a
single constraint.
%
\section{Numerical Results} \label{sec:results}
%
%
%
\subsection{Parameter Setup and Methodology}  \label{subsec:results}
In this section, we compare the performance and computation time
of the different methods to solve problem (\ref{eq:objzf}). We
used a Rayleigh fading model to generate the user channels and
assumed independent fading between users, antennas and
subcarriers. Unless noted otherwise, we used a configuration with
$M=3$ transmit antennas, $K=16$ users, $N=16$ subcarriers, and one
RT user (i.e., $D=1$). We also fixed the power constraint to
$\check{P}=20$
 and used a
large scale attenuation of 0 dB for all users. The user weights in
(\ref{eq:objzf}) were set to $c_k=1$ for all users.

We compared the performance of the different methods for various scenarios where we increased the resource requirements for the RT users until the
solution was unfeasible (i.e., the minimum rate requirements can no longer be respected for all RT users). In the first scenario, we increased the minimum
rate $\check{d}_k$ for a single user with RT QoS requirements. In the second scenario
we fixed the minimum rate but increased the large scale attenuation of the single RT user to
study the case where RT users travel away from the BS while
still having minimum rate requirements.  In the last scenario, we increased the number of RT users with the same minimum rate requirements.

For each scenario and channel realization, the dual solution upper
bound was computed using algorithm \ref{upper-bound-alg} described
in Section~\ref{sec:comp-dual-funct}. For small system
configurations we also found the exact solution using the primal
enumeration algorithm \ref{algo:enum-alg} given in
Section~\ref{sec:finding-an-exact}. We could only find this exact
solution for small configurations because the computation time is
prohibitive for large system configurations. We also computed the
solutions given by dual-based primal feasible algorithm
\ref{algo:dualfeas} and the weight adjustment algorithm
\ref{subsubsec:ck-method-alg} described in
Section~\ref{sec:find-feas-solut}  and
\ref{sec:weight-adjustm-meth}, respectively. The provided results
are the average total rate and computation time over 100
independent channel realizations. Furthermore, whenever a result
is provided for a given configuration and algorithm, the solution
was feasible with respect to the minimum rate requirements.  We
also use the bound given by the dual optimal solution as a
reference given that the exact solution is generally not available
except for very small size cases.


%
\begin{table}
\centering
 \begin{tabular}{ | c |c | c | c |}
 \hline
Method & \multicolumn{3}{|c|}{Minimum rate (bps/Hz)} \\
\cline{2-4}
& $13.33$ &  $16.66$ &  $20$ \\
 \hline\hline
Dual-based upper bound (bps/Hz) & 49.13 & 47.12 & 40.8 \\
 \hline
Primal enum. gap (\%) &  0.57  &  0.55  & 0.10  \\
 \hline
Dual-based feas. gap (\%) &   0.57 &  0.59 &  0.04  \\
 \hline
Weight mod. gap (\%) & 0.68  &  0.71  &  0.15  \\
 \hline
 \end{tabular}
\caption  {Average performance gap against the dual optimal upper bound for small system configuration.}
 \label{tab:res-comp}
\end{table}
%
\subsection{Total Rate  Performance}
\label{subsec-rmin-results}
First we present in
Table \ref{tab:res-comp}
the average gap in percent between the three different methods to find feasible solutions against the dual optimal upper bound for a small system
configuration with  $K=4$ users and $N=2$ subcarriers.  We increased the minimum rate requirement for one RT user from 13.33 to 20 bps/Hz.
As the minimum rate increases,  the upper bound decreases as
more resources need to be assigned to the RT user until the problem
is no longer feasible.   For
this small configuration,
we see that all methods give excellent results and the duality gap is very small. Also note that due to the solver numerical accuracy and the fact that all solutions were close to each other, we even had cases where the dual-based feasible solution was better than the solution given by the primal enumeration method.

%
\begin{table}
\centering
\begin{tabular}{ | c |c | c | c | c |}
\hline
 Method & \multicolumn{3}{c|}{Minimum rate (bps/Hz)} \\
\hline
 &  80 &   100 &   120\\
\hline
\multicolumn{4}{|c|}{ Total rate gap against the upper bound (\%)} \\
\hline
Dual-based feas. &   0.24   &  0.23   &  0.21   \\
\hline
Weight mod. &   9.49  &   7.30  &  3.36    \\
\hline
\multicolumn{4}{|c|}{ Computation time (sec) } \\
\hline
Dual-based feas. &  5.30  &   5.09  &  4.22  \\
\hline
Weight mod. &   1.34  &  1.34  &  1.34  \\
\hline
\end{tabular}
\caption {Average total rate gap and computation time as a function minimum rate requirement.}
\label{tab:conf2-Rmin}
\end{table}

\begin{figure}
\centering
\includegraphics[scale=0.75]{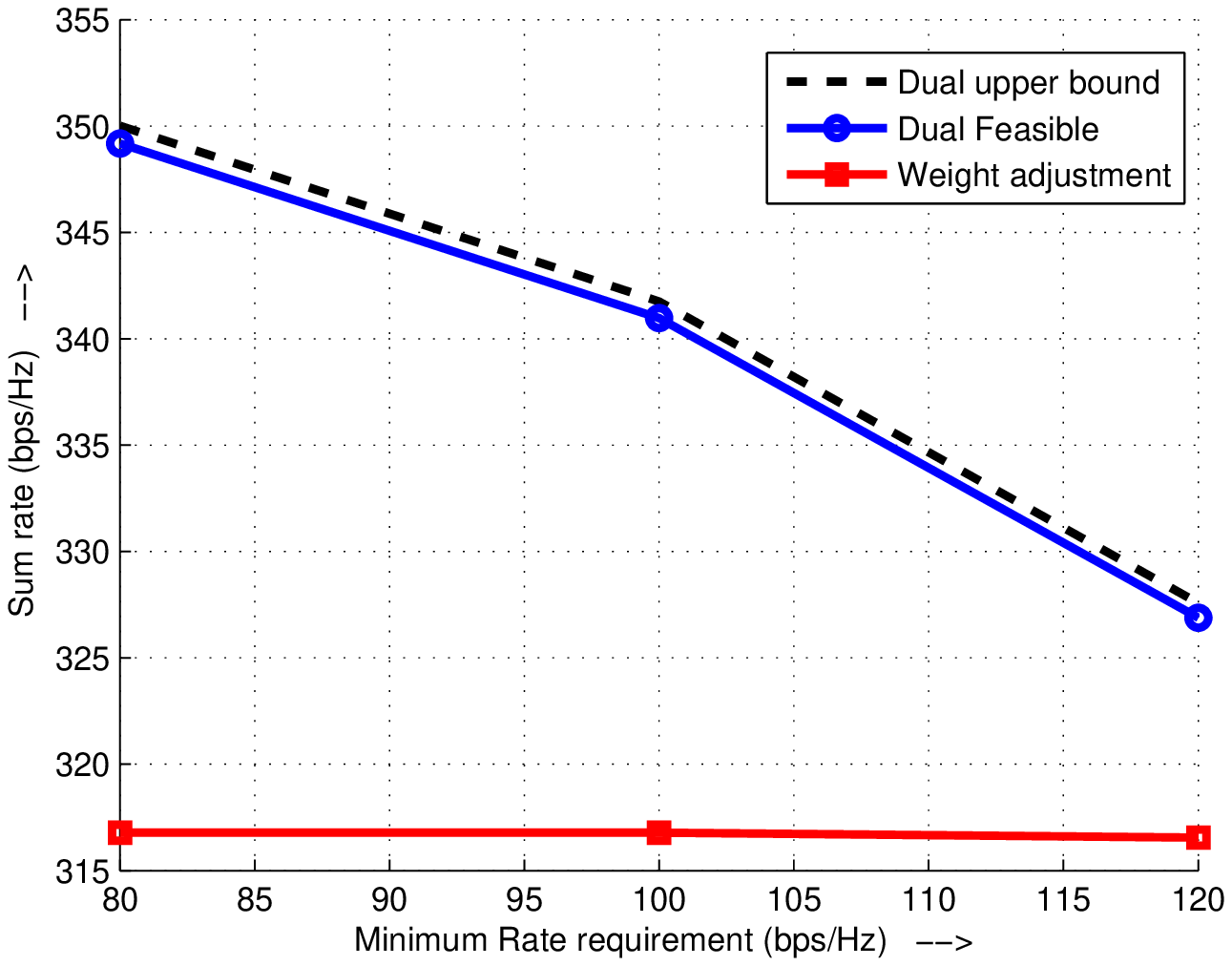}
\caption{Average total rate as a function of the minimum rate requirements} \label{fig:perf-rmin}
\end{figure}

In the remaining results we use the larger system configuration
with  $K=16$ users and $N=16$ subcarriers, where it is no longer
feasible to compute the solution using the primal enumeration
method. We present in Table \ref{tab:conf2-Rmin} the difference in
percentage between the solutions of the dual-based feasible
algorithm and the weight adjustment algorithm and the dual upper
bound. The dual-based feasible algorithm provides a solution
within $0.25$\% of the dual upper bound. The weight adjustment
method, on the other hand, is much worse and the difference can be
almost 10\%. This is due to the fact that the solution found by
the weight adjustment algorithm does not change much when the
minimum rate is increased, as can be seen from Figure
\ref{fig:perf-rmin} which shows the sum rate achieved by the
dual-based feasible algorithm and the weight modification method
against the minimum rate requirement.
%

\begin{figure}
\centering
\includegraphics[scale=0.75]{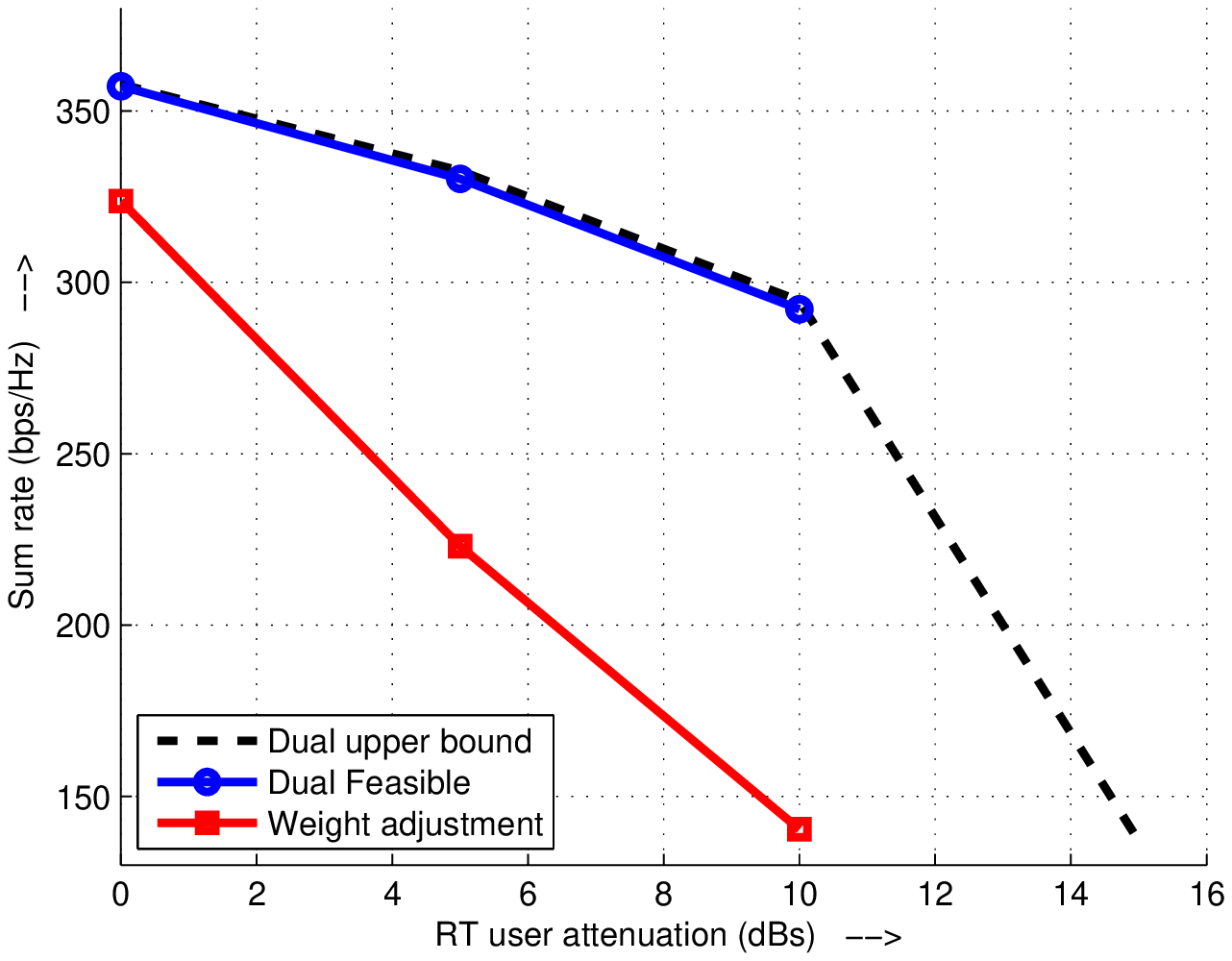}
\caption{Average total rate  as a function of RT user large scale channel attenuation.} \label{fig:att-chan}
\end{figure}

\begin{table}
\centering
\begin{tabular}{ | c |c | c | c |}
\hline
Method & \multicolumn{3}{c|}{RT user attenuation (dB) } \\
\hline
& 0 &  5 &  10 \\
\hline
\multicolumn{4}{|c|}{ Total rate gap against the upper bound (\%) } \\
\hline
Dual-based feas. &   0.16   &  0.70   &  0.82   \\
\hline
Weight mod. &   9.53  &  32.95  &  52.35    \\
\hline
\multicolumn{4}{|c|}{ Computation time (sec)} \\
\hline
Dual-based feas. &   4.73  &  5.38  &  4.58  \\
\hline
Weight mod. &  1.48 &  1.35 &  1.55  \\
\hline
\end{tabular}\caption {Average total rate gap and computation time as a function of RT user large scale channel attenuation.}
 \label{tab:conf2-SNR}
\end{table}

%

Figure~\ref{fig:att-chan} shows the average total rate when the large scale channel attenuation of the RT user varies from 0 to 15 dB. As the user moves away from the BS, the RA algorithm
dedicates more resources to this RT user until the problem is
unfeasible. The results show that for all SNR,  the
dual based method provides a solution much closer to the upper bound
than the weight adjustment method.  Table
\ref{tab:conf2-SNR} shows  the error in percentage between the
objective and the upper bound.  For  an  attenuation  of
15 dB, neither method is able to find a feasible
solution.
\begin{table}
\centering
\begin{tabular}{ | c |c | c | c | c |  c | c | c | } \hline
Method & \multicolumn{7}{|c|}{ Number of RT users} \\
\hline
& 1 & 2 & 3 & 4 & 5 & 6 & 7 \\
\hline
& \multicolumn{7}{|c|}{  Total rate gap against the upper bound (\%) } \\
\hline
Dual Feas. &   0.16   & 0.61   &  2.09   &  2.41   &  3.52   &  3.20   &  3.43   \\
\hline
Weight mod. &  3.5    &  3.5   &  6.52    &  13.86    &  22.71    &  - & - \\
\hline
& \multicolumn{7}{|c|}{Computation time (sec) } \\
\hline
Dual Feas. &   4.73 & 7.13 & 9.68 & 10.80 & 13.29 & 15.25 & 18.45  \\
\hline
Weight mod. &  1.48 & 1.50 & 1.60 & 1.69 & 2.06 & - & - \\
\hline
\end{tabular}\caption {Average total rate gap and computation time as a function of the number of RT users}
 \label{tab:conf2-Lusers}
\end{table}
%

\begin{figure}
\centering
\includegraphics[scale=0.75]{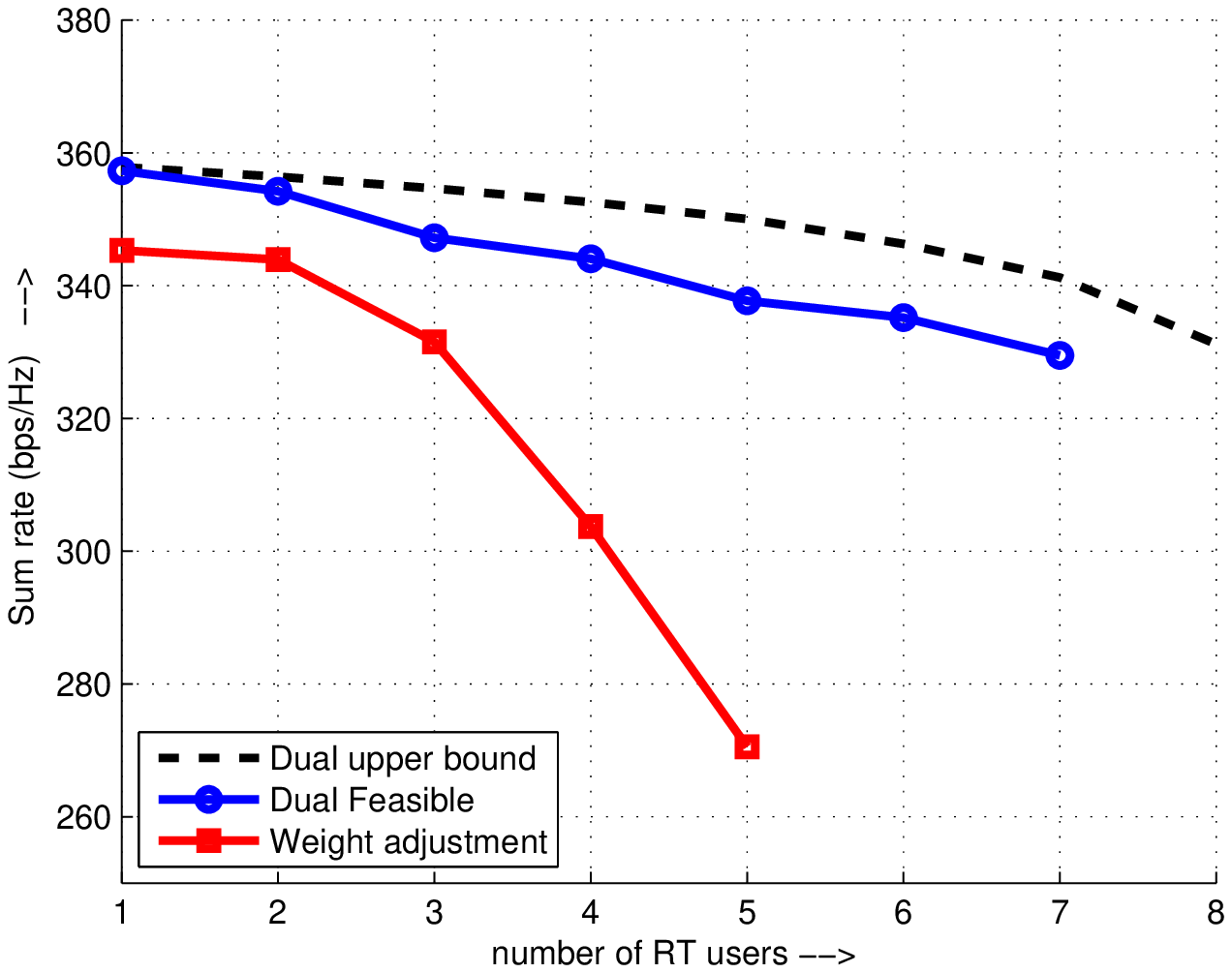}
\caption{Average total rate  as a function of the number of RT users.}
\label{fig:perf-Lusers}
\end{figure}
%
%

Finally, Figure \ref{fig:perf-Lusers} shows the optimal dual bound
and the solution given by the dual-based feasible and weight
adjustment methods as a function of the number of RT users. Table
\ref{tab:conf2-Lusers} lists the performance gap against the dual
bound   in percentage.   The dual feasible method exhibits a much
lower performance gap than the weight adjustment method for all
values of the number of RT users.  Moreover, we can see that the
weight adjustment method performance quickly degrades when the
number of RT users increases. It is not even able to find feasible
points when the number of RT users is $6$ or $7$ while the
dual-based feasible algorithm provides solutions within $3.52 \%$
of the upper bound.

We can observe from all the results presented in this section
(similar trends were also observed for other configurations that
we studied) that the difference between the dual-based upper bound
and the dual-based feasible solution is very small. This is an
indication that the duality gap is very small and that it is
possible to find feasible solutions close to the optimal, albeit
with an off-line algorithm. Furthermore, the dual-based feasible
solution is always better than the weight modification method and
this difference becomes more significant as the resource
requirements to meet the RT users needs increase. This shows that
the weight modification method should be used carefully for RA in
OFDMA-SDMA systems with RT users and that more efficient
heuristics should be developed to approach the performance of the
dual-based feasible solution.

\subsection{Computation Time} \label{subsec-complexity-results}
We compare the computation time of the algorithms based on the
time required by an Intel dual-core 660 CPU to solve the problem.
The primal enumeration method is not included in the comparison
due to its computation inefficiency. The average computation time
to find a solution for the three scenarios (i.e. variation of
minimum rate requirement, large scale channel attenuation and
number of RT users) is presented in Tables~\ref{tab:conf2-Rmin} to
\ref{tab:conf2-Lusers}.

In Table \ref{tab:conf2-Rmin}, where the minimum rate constraints
are varied, the computation time remains constant for the weight
adjustment method, requiring an average of two main iterations of
algorithm \ref{subsubsec:ck-method-alg}. The dual-based method
takes on average $3.63$ times more time than the weight adjustment
method.  However, the total rate that it produces  is much closer
to the upper bound. We have similar results in Table
\ref{tab:conf2-SNR},  where we varied the SNR of the RT user. The
dual-based method produces a higher total sum rate than that of
the weight adjustment method at the expense of a larger CPU time.

Table \ref{tab:conf2-Lusers} shows the computation time as a
function of the number of RT users.  For the weight adjustment
method the computation time is almost constant and slightly
increases until the number of RT users is five, after which it
cannot find feasible solutions. In contrast, the dual-based method
computation time grows steadily with the number of RT users and it
is approximately five times higher.

This comparison highlights the main difference between these two
methods: While the weight adjustment method performs several
iterations solving a problem with a single constraint, the dual
method takes into account the rate constraints  so that the computation
time increases with the number of RT users. However, the
performance gap against the dual upper bound of the dual method is
very low and approximately constant.  In contrast, the performance
gap of the weight adjustment method is larger and increases with
the number of RT users.  Moreover, the dual method can find
feasible solutions for cases where the weight adjustment method cannot.
\section{Conclusion} \label{sec:conclusions}
In this work, we  proposed a method to compute the beamforming
vectors and the user selection in an OFDMA-SDMA MISO system with
minimum rates for some RT users. We used  a Lagrangian
relaxation  of the power and rate constraints and solved the dual
problem using a subgradient algorithm.  The Lagrange
decomposition yields sub-problems separated per subcarrier,
SDMA sets and users which substantially reduces the computational
complexity. We also used a closed-form expression of
the beamforming subproblem  based on a pseudo-inverse condition on the
beamforming vectors. The dual function is expressed in terms of
the dual variables and  the dual optimum is found using a
subgradient algorithm.  The dual optimum can then be used as a
benchmark to compare against other solution methods and
heuristics.

We then proposed, in addition to the complete enumeration approach
which is not computationally practical for normal size problems,
two algorithms to find feasible solution to the RA problem with
minimum rate requirements. The first algorithm starts from the
dual-based optimum solution and finds a feasible solution by
searching along the rate requirement dual variables, while the
second uses weight adjustments in the objective function to
achieve the required rates. Our results show that the dual-based
method provides solutions much closer to the upper bound than the
weight adjustment method. The difference is more significant when
the SNR of the RT users is low or the number of RT users is high.
However, the computational load of the dual-based method increases
when the number of users with minimum rate requirements increases,
whereas it is almost constant for the weight adjustment method.
This reflects the trade-off between the two methods. The weight
adjustment method is computationally more efficient but the
solutions provided by the proposed dual-based method are much
closer to the optimal. This indicates that there is an advantage
when including the minimum rate constraints in the resource
allocation problem. In addition, the weight adjustment method
requires many time slots to adjust the weights and schedule real
time users.  Our method explicitly includes the minimum rate
constraints which allows RT users to be scheduled in the current
slot, which decreases the average packet delay.

To implement the RA algorithm in real time, we still need to
design fast methods to reduce the number of SDMA sets to be
searched. The design of these heuristic algorithms is outside the
scope of this paper but it is part of our current efforts.
Finally, the upper bound given by the dual function minimization
provides a very useful benchmark to compare the performance of
these heuristics and the dual-based algorithm can also guide the
design of efficient novel heuristics.

\section*{Acknowledgments}
This research project was  partially supported by NSERC grant CRDPJ 335934-06.

\pagebreak
\singlespacing
%
\bibliographystyle{ieeetran}
\bibliography{journal_paper}
\end{document}